\documentclass[
reprint,
amsmath,
amssymb,
aip,
pop
]{revtex4-2}

\usepackage[USenglish]{babel}
\usepackage{graphicx}
\usepackage{booktabs}
\usepackage{url}
\usepackage{tikz,circuitikz}
\usetikzlibrary{decorations.pathreplacing,arrows.meta,calc,patterns,positioning}

\begin{document}
	
\title{A Minimal Self-Consistent Model for the Nonlinear Dynamics of Asymmetric Capacitively Coupled Radio-Frequency Plasmas}
\author{Thomas Mussenbrock}
\email{Thomas.Mussenbrock@rub.de}
\affiliation{Chair of Applied Electrodynamics and Plasma Technology, Faculty of Electrical Engineering and Information Technology, Ruhr University Bochum, 44780 Bochum, Germany}
\date{\today}
	
\begin{abstract}
	We develop a minimal self-consistent lumped-element model for the nonlinear radio-frequency dynamics of geometrically asymmetric capacitively coupled plasmas. The fast electrical response is described by three dynamical variables: the powered-sheath charge, the blocking-capacitor voltage, and the discharge current. The periodic RF solution is coupled to stationary particle and energy balances, which determine the electron temperature and plasma density from the operating conditions and the absorbed power. For an argon discharge, the model produces nonlinear current oscillations and a broad higher-harmonic response associated with the plasma series resonance. Variation of the RF voltage leads to correlated changes in the harmonic spectrum, absorbed power, and plasma density. Linearization about the periodic state relates the high-frequency response to the time-dependent differential sheath elastance. A moment-based effective elastance provides a compact estimate of the corresponding characteristic PSR frequency. The model thereby connects nonlinear RF dynamics and plasma sustainment within a minimal self-consistent description.
\end{abstract}
	
\maketitle	
	
\section{Introduction}

Capacitively coupled radio-frequency (CCRF) discharges are widely used for plasma processing because the applied voltage provides direct control of particle and energy transport to material surfaces \cite{LiebermanLichtenberg2005}. Their electrical response is intrinsically nonlinear. A substantial fraction of the applied RF voltage is supported by the plasma boundary sheaths, whose charge-voltage characteristics are nonlinear, whereas the quasineutral plasma bulk exhibits an inertial electron response. The interaction of sheath charging with the bulk electron dynamics determines the electrical impedance of the discharge and contributes to harmonic generation, electron heating, and power absorption.

Electron interaction with oscillating RF sheaths has been studied for several decades. Godyak interpreted collisionless electron heating in terms of the statistical interaction of electrons with an oscillating plasma boundary \cite{Godyak1972}. This picture was developed further by Goedde, Lichtenberg, and Lieberman in a self-consistent stochastic-heating model \cite{GoeddeLichtenbergLieberman1988}. Wendt and Hitchon, Surendra and Graves, and Turner and Hopkins investigated electron interaction with moving RF sheaths using analytical and kinetic descriptions \cite{WendtHitchon1992,SurendraGraves1991,TurnerHopkins1992}. Related studies by Surendra and Vender and by Buddemeier, Kortshagen, and Pukropski further examined collisionless energy transfer in low-pressure capacitive discharges \cite{SurendraVender1994,BuddemeierKortshagenPukropski1995}. Later work showed that the simple picture of reflection from a rigid moving boundary is incomplete. Electron-pressure effects, the self-consistent sheath motion, and the spatial structure of the RF electric field can all contribute to the energy transfer \cite{Turner1995,AlievKaganovichSchlueter1997,LiebermanGodyak1998,GozadinosEtAl2001PSST,GozadinosTurnerVender2001PRL,Kaganovich2002PRL}. The relation between the hard-wall approximation and the self-consistent discharge dynamics was further clarified by Kawamura, Lieberman, and Lichtenberg \cite{KawamuraLiebermanLichtenberg2006}; a broader discussion of collisionless heating mechanisms is given by Turner \cite{Turner2009Review}.

The capacitive sheath response and the inertia of the plasma electrons also give rise to a natural series resonance of the discharge. Godyak and Popov investigated resonant RF discharges experimentally at an early stage \cite{GodyakPopov1979}. Plasma-sheath resonances in parallel-plate reactors were later identified experimentally by Annaratone, Allen, Ku, and co-workers and interpreted in terms of equivalent circuits containing capacitive plasma sheaths and an inductive plasma bulk \cite{AnnaratoneAllenKu1995,KuAnnaratoneAllen1998a,KuAnnaratoneAllen1998b}. Related resonant states were studied numerically and experimentally by Cooperberg and Birdsall and by Qiu, Bowers, and Birdsall \cite{CooperbergBirdsall1998,QiuBowersBirdsall2003}.

In a nonlinear discharge, the sheath charge-voltage relation generates higher harmonics of the applied excitation \cite{Klick1996}. These harmonics can couple to the inertial bulk response and excite the plasma series resonance (PSR) even when the applied RF frequency lies well below the characteristic resonance frequency. Klick, Rehak, and Kammeyer exploited the resulting nonlinear current response for self-excited electron resonance spectroscopy \cite{KlickRehakKammeyer1997}. Nonlinear equivalent-circuit and global descriptions subsequently established the connection between harmonic generation and self-excited PSR oscillations in low-pressure CCRF discharges \cite{MussenbrockBrinkmann2006APL,MussenbrockZieglerBrinkmann2006POP,CzarnetzkiEtAl2006POP,Mussenbrock2008PRL,Lieberman2008POP}. Experiments by Czarnetzki and co-workers confirmed high-frequency current oscillations associated with the nonlinear sheath--bulk interaction and their connection to electron heating \cite{SchulzeEtAl2007JPCS}. Particle simulations later showed that PSR excitation is not restricted to geometrically asymmetric discharges and may also arise from electrical asymmetry and higher-order nonlinearities \cite{DonkoEtAl2009APL,SchuengelEtAl2015POP}. Geometrical asymmetry nevertheless remains important for the nonlinear electron dynamics and power absorption \cite{NoesgesMussenbrock2025POP}.

Reduced lumped-element models are useful in this context because they retain the essential sheath--bulk interaction with only a small number of physical degrees of freedom. Their applicability is restricted to electrically small discharges for which electromagnetic propagation, standing-wave effects, and skin effects remain weak \cite{LiebermanBoothChabert2002,MussenbrockBrinkmann2007PSST,Chabert2007,LeeGravesLieberman2008,MussenbrockHemkeZiegler2008PSST}. Within this regime, nonlinear sheath charging, electron inertia, dissipation, and the external RF circuit can be treated without resolving the spatial RF field.

Existing nonlinear circuit models already contain the main ingredients of the fast electrical dynamics. The macroscopic plasma state, however, is often prescribed or determined separately from the nonlinear RF solution. The feedback between absorbed RF power, plasma density, and the parameters controlling the electrical response is then not retained within the reduced model. A second difficulty concerns the definition of a characteristic PSR frequency. Since the differential sheath capacitance varies strongly over an RF period, the nonlinear periodic state cannot in general be represented by a single fixed sheath capacitance and a unique linear resonance frequency.

In the present work, the nonlinear RF dynamics and the stationary plasma state are coupled within a minimal self-consistent model. The fast subsystem contains only three dynamical variables: the powered-sheath charge, the blocking-capacitor voltage, and the discharge current. The nonlinear sheath generates higher harmonics, while electron inertia in the plasma bulk provides the inductive response required for PSR excitation. Geometrical asymmetry is represented by different effective powered- and grounded-electrode areas. The periodic RF solution is coupled to stationary particle and energy balances, so that the electron temperature and plasma density follow from the operating conditions and the cycle-averaged absorbed power.

We further examine how the time-dependent differential sheath elastance determines the characteristic high-frequency response of the periodic state. A moment-based effective elastance is introduced as a reduced measure of this time-dependent quantity and is compared with the harmonic response obtained from the nonlinear RF solution. This connects the stationary plasma state, the nonlinear electrical dynamics, and the characteristic PSR scale within the same reduced model.

\section{Minimal lumped-element model of the RF discharge}

\subsection{Scale separation and model assumptions}

We consider a low-pressure capacitively coupled radio-frequency discharge in which the characteristic electromagnetic and plasma scales permit a lumped-element description. The relevant ordering of spatial scales is
\begin{gather}
	\lambda_{\mathrm D}
	\ll
	l_{\mathrm s}
	<
	d_{\mathrm{gap}}
	\ll
	\lambda_{\mathrm{EM}},
\end{gather}
where $\lambda_{\mathrm D}$ is the Debye length, $l_{\mathrm s}$ a characteristic sheath thickness, $d_{\mathrm{gap}}$ the axial electrode separation, and $\lambda_{\mathrm{EM}}$ the electromagnetic wavelength associated with the applied RF excitation. The condition $\lambda_{\mathrm D}\ll l_{\mathrm s}$ permits the non-neutral boundary sheath to be distinguished from the quasineutral plasma bulk \cite{LiebermanLichtenberg2005,MussenbrockZieglerBrinkmann2006POP,Lieberman2008POP}. We do not require the stronger asymptotic ordering $l_{\mathrm s}\ll d_{\mathrm{gap}}$. The reduced bulk model introduced below uses $d_{\mathrm{gap}}$ as its characteristic axial length, and the validity of this approximation is checked a posteriori for the reference discharge.

The condition $
	d_{\mathrm{gap}}
	\ll
	\lambda_{\mathrm{EM}}$
allows electromagnetic propagation across the discharge to be neglected. We further restrict the model to conditions for which standing-wave and skin effects remain weak, so that the RF field can be treated within the electroquasistatic approximation. The limits of this approximation at high plasma density, large reactor dimensions, or high excitation frequency have been discussed extensively \cite{LiebermanBoothChabert2002,Chabert2007,MussenbrockBrinkmann2007PSST,LeeGravesLieberman2008,MussenbrockHemkeZiegler2008PSST}.

The characteristic plasma frequencies satisfy
\begin{gather}
	\omega_{\mathrm{pi}}
	\ll
	\omega_{\mathrm{RF}}
	\ll
	\omega_{\mathrm{pe}},
\end{gather}
where $\omega_{\mathrm{pi}}$ and $\omega_{\mathrm{pe}}$ are the ion and electron plasma frequencies. The ions therefore respond predominantly to the phase-averaged field on the RF timescale, whereas the electrons follow the instantaneous RF field. Their inertia must be retained in the quasineutral bulk because it provides the inductive response required for the plasma series resonance. For the operating conditions considered here, the characteristic high-frequency response associated with the PSR lies above the applied RF frequency and well below the electron plasma frequency.

The macroscopic plasma state evolves on a timescale much longer than one RF period. The electron density $n_{\mathrm e}$, electron temperature $T_{\mathrm e}$, and the kinetic coefficients entering the RF model are therefore taken to be constant during one RF cycle. Their stationary values are determined from the zero-dimensional plasma closure introduced below.

Within the electroquasistatic approximation, charge conservation requires the same total current, including displacement current, to pass through the successive regions of the discharge. The powered-electrode sheath is represented by a nonlinear charge-storage element, while the quasineutral bulk is represented by an inductive-resistive element accounting for electron inertia and dissipation. The grounded-electrode sheath is retained in reduced form as described below. This decomposition leads to the equivalent circuit introduced in the following subsection.

\subsection{Equivalent circuit representation}
	
\begin{figure}[t]
		\centering
				    \resizebox{\columnwidth}{!}{%
			
		\begin{tikzpicture}[
			>=Latex,
			line width=0.8pt,
			font=\small
			]
			
			% ------------------------------------------------------------
			% Geometry
			% ------------------------------------------------------------
			
			\def\rE{1.15}
			\def\rG{2.35}
			\def\h{6.2}
			
			% Electrode vertical positions
			\coordinate (ET) at (0,0);
			\coordinate (EG) at (0,-\h);
			
			% ------------------------------------------------------------
			% Plasma bulk: conical frustum
			% ------------------------------------------------------------
			
			\fill[blue!5]
			(-\rE,0)
			--
			(-\rG,-\h)
			arc[start angle=180,end angle=360,x radius=\rG,y radius=0.36]
			--
			(\rE,0)
			arc[start angle=0,end angle=180,x radius=\rE,y radius=0.18]
			-- cycle;
			
			% Side walls
			\draw
			(-\rE,0)
			--
			(-\rG,-\h);
			
			\draw
			(\rE,0)
			--
			(\rG,-\h);
			
			% ------------------------------------------------------------
			% Powered electrode
			% ------------------------------------------------------------
			
			\fill[
			red!25,
			pattern=north east lines,
			pattern color=red!65
			]
			(0,0)
			ellipse[x radius=\rE,y radius=0.18];
			
			\draw
			(0,0)
			ellipse[x radius=\rE,y radius=0.18];
			
			% ------------------------------------------------------------
			% Grounded electrode
			% ------------------------------------------------------------
			
			\fill[
			blue!18,
			pattern=north east lines,
			pattern color=blue!45
			]
			(0,-\h)
			ellipse[x radius=\rG,y radius=0.36];
			
			\draw
			(0,-\h)
			ellipse[x radius=\rG,y radius=0.36];
			
			% ------------------------------------------------------------
			% RF source
			% ------------------------------------------------------------
			
			\draw (0,0.18) -- (0,0.95);
			
			\draw (0,1.35) circle[radius=0.40];
			
			\node at (0,1.35) {$\sim$};
			
			\draw (0,1.75) -- (0,2.05);
			
			% ------------------------------------------------------------
			% Ground
			% ------------------------------------------------------------
			
			\draw (0,-\h-0.36) -- (0,-\h-0.85);
			
			\draw (-0.32,-\h-0.85) -- (0.32,-\h-0.85);
			\draw (-0.22,-\h-0.98) -- (0.22,-\h-0.98);
			\draw (-0.12,-\h-1.11) -- (0.12,-\h-1.11);
			
			% ------------------------------------------------------------
			% Axial coordinate
			% ------------------------------------------------------------
			
			\draw[->]
			(0,-0.35)
			--
			(0,-1.75);
			
			\node[right] at (0,-1.05) {$z$};
			
			% ------------------------------------------------------------
			% Radius r_E
			% ------------------------------------------------------------
			
			\draw[<->]
			(0,0.45)
			--
			(\rE,0.45);
			
			\node[above] at ({0.5*\rE},0.45) {$r_{\mathrm E}$};
			
			% ------------------------------------------------------------
			% Radius r_G
			% ------------------------------------------------------------
			
			\draw[<->]
			(0,-\h+0.55)
			--
			(\rG,-\h+0.55);
			
			\node[above] at ({0.5*\rG},-\h+0.55) {$r_{\mathrm G}$};
			
			% ------------------------------------------------------------
			% Local cross section
			% ------------------------------------------------------------
			
			\def\zloc{-4.25}
			\pgfmathsetmacro{\rloc}{
				\rE + (\rG-\rE)*abs(\zloc)/\h
			}
			
			\draw[
			blue,
			dashed
			]
			(0,\zloc)
			ellipse[
			x radius=\rloc,
			y radius=0.25
			];
			
			\draw[
			blue,
			<->,
			line width=0.7pt
			]
			(0,\zloc)
			--
			(\rloc,\zloc);
			
			\node[
			blue,
			above
			] at ({0.5*\rloc},0.95*\zloc)
			{$r(z)$};
			
			% ------------------------------------------------------------
			% Gap distance
			% ------------------------------------------------------------
			
			\draw
			(\rG+0.55,0)
			--
			(\rG+0.95,0);
			
			\draw
			(\rG+0.55,-\h)
			--
			(\rG+0.95,-\h);
			
			\draw[<->]
			(\rG+0.80,-0.05)
			--
			(\rG+0.80,-\h+0.05);
			
			\node[
			right
			] at (\rG+0.80,-0.5*\h)
			{$d_{\mathrm{gap}}$};
			
			% ------------------------------------------------------------
			% Coordinate labels
			% ------------------------------------------------------------
			
			\node[
			right
			] at (\rE+0.30,0)
			{$z=0$};
			
			\node[
			right
			] at (\rG+1.0,-\h)
			{$z=d_{\mathrm{gap}}$};
			
			% ------------------------------------------------------------
			% Labels
			% ------------------------------------------------------------
			
			\node[
			red!75!black,
			align=center,
			left
			] at (-\rE-0.0,0.20)
			{
				Powered electrode\\
				$A_{\mathrm E}=\pi r_{\mathrm E}^2$
			};
			
			\node[
			blue!70!black,
			align=center,
			right
			] at (\rG-1.3,-\h-0.9)
			{
				Grounded electrode\\
				$A_{\mathrm G}=\pi r_{\mathrm G}^2$
			};
			
			\node[
			blue!70!black,
			align=center
			] at (0,-2.8)
			{
				Plasma bulk\\
				(quasineutral)
			};
			
			% ------------------------------------------------------------
			% Radius law
			% ------------------------------------------------------------
			
			\node[
			align=left,
			anchor=west
			] at (-4.8,-1.5)
			{
				$r(z)
				=
				r_{\mathrm E}
				+
				\dfrac{
					r_{\mathrm G}-r_{\mathrm E}
				}{
					d_{\mathrm{gap}}
				}z$
			};
			
			\node[
			align=left,
			anchor=west
			] at (-4.8,-2.25)
			{
				$A(z)=\pi r^2(z)$
			};
			
		\end{tikzpicture}
	}
\caption{
	Equivalent conical-frustum geometry used for the asymmetric RF discharge. The frustum represents an effective current channel with end areas $A_{\mathrm E}$ and $A_{\mathrm G}$ and axial separation $d_{\mathrm{gap}}$, rather than the detailed reactor geometry.
}
		\label{fig:equivalent_geometry}
	\end{figure}
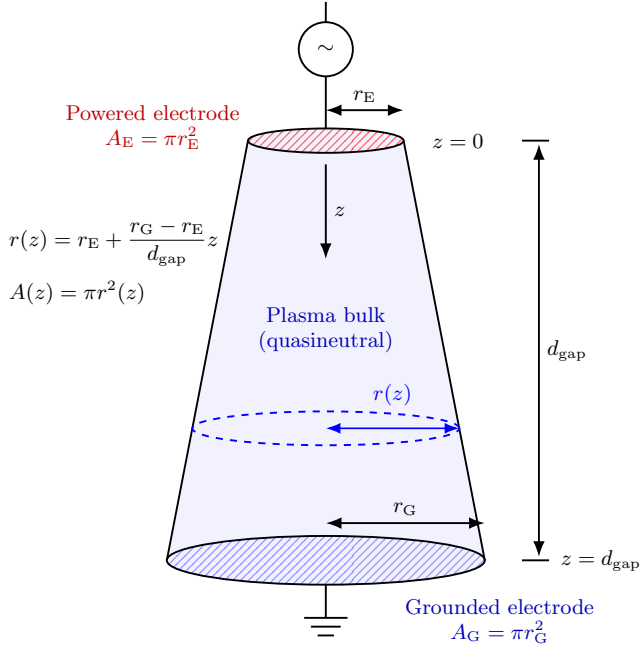

The reactor geometry is represented by an equivalent axially expanding current channel with the shape of a conical frustum, as shown in Fig.~\ref{fig:equivalent_geometry}. Its end areas correspond to the effective powered- and grounded-electrode areas $A_{\mathrm E}$ and $A_{\mathrm G}$, respectively, while $d_{\mathrm{gap}}$ denotes their axial separation. This geometry is not intended to reproduce the detailed reactor shape. It retains only the geometrical asymmetry and the corresponding variation of current density along the discharge.

Introducing the axial coordinate
\begin{gather}
	0
	\leq
	z
	\leq
	d_{\mathrm{gap}},
\end{gather}
the radius of the equivalent current channel is taken to vary linearly,
\begin{gather}
	r(z)
	=
	r_{\mathrm E}
	+
	\frac{r_{\mathrm G}-r_{\mathrm E}}
	{d_{\mathrm{gap}}}
	z ,
	\label{eq:frustum_radius}
\end{gather}
with
\begin{gather}
	A(z)
	=
	\pi r^2(z).
	\label{eq:frustum_area}
\end{gather}
The end areas are
\begin{align}
	A_{\mathrm E}
	&=
	\pi r_{\mathrm E}^2,
	\\
	A_{\mathrm G}
	&=
	\pi r_{\mathrm G}^2,
\end{align}
and the geometrical asymmetry is characterized by
\begin{gather}
	\mathcal{A}
	=
	\frac{A_{\mathrm G}}{A_{\mathrm E}}
	=
	\left(
	\frac{r_{\mathrm G}}{r_{\mathrm E}}
	\right)^2
	>
	1.
	\label{eq:geometrical_asymmetry}
\end{gather}

The discharge is divided into the powered-electrode sheath, the quasineutral plasma bulk, and the grounded-electrode sheath. The same total current $I(t)$, including displacement current, passes through all three regions.

\paragraph{Powered-electrode sheath.}

The powered sheath is described by the nonlinear charge-voltage relation introduced by Metze et al.~\cite{Metze1986} and used subsequently in reduced models of capacitively coupled RF discharges \cite{MussenbrockBrinkmann2006APL,MussenbrockZieglerBrinkmann2006POP,Mussenbrock2008PRL,Lieberman2008POP}.

The sheath voltage is defined as
\begin{gather}
	V_{\mathrm S}
	=
	\phi_{\mathrm p}
	-
	\phi_{\mathrm E}
	>
	0,
	\label{eq:sheath_voltage_definition}
\end{gather}
where $\phi_{\mathrm p}$ and $\phi_{\mathrm E}$ are the plasma and powered-electrode potentials.

Let $Q_{\mathrm S}\geq0$ denote the positive space charge in the sheath. Approximating the ion density by the sheath-edge density $n_{\mathrm s}$ gives
\begin{gather}
	Q_{\mathrm S}
	=
	e n_{\mathrm s} A_{\mathrm E}s,
	\label{eq:sheath_charge_width}
\end{gather}
where $s$ is the instantaneous sheath width. The corresponding charge-voltage relation is
\begin{gather}
	V_{\mathrm S}(Q_{\mathrm S})
	=
	\frac{
		Q_{\mathrm S}^{2}
	}{
		2e\varepsilon_0 n_{\mathrm s}A_{\mathrm E}^{2}
	}.
	\label{eq:sheath_voltage}
\end{gather}
Equivalently,
\begin{gather}
	s(t)
	=
	\sqrt{
		\frac{
			2\varepsilon_0 V_{\mathrm S}(t)
		}{
			e n_{\mathrm s}
		}
	}.
	\label{eq:sheath_width}
\end{gather}
This relation will later be used to assess the sheath length scale for the reference discharge.

The differential sheath elastance is
\begin{gather}
	S_{\mathrm S}(Q_{\mathrm S})
	=
	\frac{\partial V_{\mathrm S}}
	{\partial Q_{\mathrm S}}
	=
	\frac{
		Q_{\mathrm S}
	}{
		e\varepsilon_0 n_{\mathrm s}A_{\mathrm E}^{2}
	}
	=
	\frac{1}
	{C_{\mathrm S}(Q_{\mathrm S})}.
	\label{eq:sheath_elastance}
\end{gather}
Since $Q_{\mathrm S}$ varies during the RF period, both $S_{\mathrm S}$ and the differential capacitance $C_{\mathrm S}$ are time dependent. This time dependence becomes important for the high-frequency response discussed below.

The model is restricted to
\begin{gather}
	Q_{\mathrm S}
	\geq
	0,
	\label{eq:sheath_charge_constraint}
\end{gather}
with $Q_{\mathrm S}=0$ corresponding to sheath collapse. Negative sheath charge lies outside the present sheath model and is excluded numerically.

Particle currents through the sheath are included in addition to its displacement current. The ion current toward the powered electrode is approximated by the Bohm flux,
\begin{gather}
	I_{\mathrm i}
	=
	e n_{\mathrm s}v_{\mathrm B}A_{\mathrm E},
	\label{eq:ion_current}
\end{gather}
with
\begin{gather}
	v_{\mathrm B}
	=
	\sqrt{
		\frac{eT_{\mathrm e}}{m_{\mathrm i}}
	},
	\label{eq:bohm_velocity_rf}
\end{gather}
while the electron current is
\begin{gather}
	I_{\mathrm e}(V_{\mathrm S})
	=
	I_{\mathrm e0}
	\exp\left(
	-\frac{V_{\mathrm S}}{T_{\mathrm e}}
	\right),
	\label{eq:electron_current}
\end{gather}
with
\begin{gather}
	I_{\mathrm e0}
	=
	\frac{1}{4}
	e n_{\mathrm s}\bar{v}_{\mathrm e}A_{\mathrm E},
	\label{eq:electron_saturation_current}
\end{gather}
and
\begin{gather}
	\bar{v}_{\mathrm e}
	=
	\sqrt{
		\frac{8eT_{\mathrm e}}
		{\pi m_{\mathrm e}}
	}.
	\label{eq:mean_electron_speed}
\end{gather}

\paragraph{Grounded-electrode sheath.}

Because the same total current passes through both electrode surfaces,
\begin{gather}
	j_{\mathrm E}(t)
	=
	\frac{I(t)}{A_{\mathrm E}},
	\label{eq:powered_current_density}
\end{gather}
and
\begin{gather}
	j_{\mathrm G}(t)
	=
	\frac{I(t)}{A_{\mathrm G}}
	=
	\frac{j_{\mathrm E}(t)}{\mathcal A}.
	\label{eq:grounded_current_density}
\end{gather}
For $\mathcal A>1$, the current-density modulation at the grounded surface is smaller than at the powered electrode. To retain a three-variable RF model, we therefore neglect the RF modulation of the grounded sheath. This is a deliberate reduction of the model, not an asymptotic consequence of geometrical asymmetry.

Earlier reduced models neglected the grounded-sheath impedance altogether \cite{Mussenbrock2008PRL,Lieberman2008POP,ZieglerMussenbrockBrinkmann2009POP}. Here, its stationary voltage drop is retained,
\begin{gather}
	V_{\mathrm G}(t)
	\simeq
	V_{\mathrm{fl}}
	=
	\mathrm{const}.
	\label{eq:grounded_sheath_voltage}
\end{gather}
The floating potential follows from equality of the stationary ion and electron fluxes,
\begin{gather}
	j_{\mathrm i,G}
	=
	j_{\mathrm e,G},
	\label{eq:floating_flux_balance}
\end{gather}
with
\begin{gather}
	j_{\mathrm i,G}
	=
	e n_{\mathrm s}v_{\mathrm B},
	\label{eq:grounded_ion_flux}
\end{gather}
and
\begin{gather}
	j_{\mathrm e,G}
	=
	\frac{1}{4}
	e n_{\mathrm s}\bar{v}_{\mathrm e}
	\exp\left(
	-\frac{V_{\mathrm{fl}}}{T_{\mathrm e}}
	\right).
	\label{eq:grounded_electron_flux}
\end{gather}
Hence,
\begin{gather}
	V_{\mathrm{fl}}
	=
	\frac{T_{\mathrm e}}{2}
	\ln\left(
	\frac{m_{\mathrm i}}
	{2\pi m_{\mathrm e}}
	\right).
	\label{eq:floating_potential}
\end{gather}
Here, $T_{\mathrm e}$ is expressed in electronvolts and $V_{\mathrm{fl}}$ in volts \cite{LiebermanLichtenberg2005}.

The grounded sheath therefore contributes a stationary voltage drop without adding another RF degree of freedom.

\paragraph{Plasma bulk.}

The quasineutral plasma bulk is described by the electron momentum balance. Ion motion is neglected on the RF timescale, and explicit electron-pressure-gradient terms are omitted. The electron current density then obeys
\begin{gather}
	\frac{\partial \vec{j}}{\partial t}
	=
	\frac{e^2 n_{\mathrm e}}{m_{\mathrm e}}\vec{E}
	-
	\nu_{\mathrm{eff}}\vec{j},
	\label{eq:bulk_ohm}
\end{gather}
where $\nu_{\mathrm{eff}}$ is the effective damping frequency.

Current continuity in the equivalent channel gives
\begin{gather}
	j(z,t)
	=
	\frac{I(t)}{A(z)}.
	\label{eq:local_current_density}
\end{gather}
Assuming $n_{\mathrm e}$ and $\nu_{\mathrm{eff}}$ to be spatially uniform,
\begin{gather}
	E(z,t)
	=
	\frac{m_{\mathrm e}}
	{e^2 n_{\mathrm e} A(z)}
	\left[
	\frac{dI}{dt}
	+
	\nu_{\mathrm{eff}} I
	\right].
	\label{eq:local_bulk_field}
\end{gather}
Integration along the equivalent current channel yields
\begin{gather}
	V_{\mathrm p}(t)
	=
	L_{\mathrm p}
	\frac{dI}{dt}
	+
	R_{\mathrm p}I(t),
	\label{eq:bulk_voltage}
\end{gather}
where
\begin{gather}
	L_{\mathrm p}
	=
	\frac{m_{\mathrm e}}
	{e^2 n_{\mathrm e}}
	\int_0^{d_{\mathrm{gap}}}
	\frac{dz}{A(z)}
	\label{eq:plasma_inductance_integral}
\end{gather}
and
\begin{gather}
	R_{\mathrm p}
	=
	\nu_{\mathrm{eff}}L_{\mathrm p}.
	\label{eq:plasma_resistance}
\end{gather}
For the conical-frustum geometry,
\begin{gather}
	\int_0^{d_{\mathrm{gap}}}
	\frac{dz}{A(z)}
	=
	\frac{d_{\mathrm{gap}}}
	{\sqrt{A_{\mathrm E}A_{\mathrm G}}},
\end{gather}
so that
\begin{gather}
	L_{\mathrm p}
	=
	\frac{
		m_{\mathrm e}d_{\mathrm{gap}}
	}{
		e^2 n_{\mathrm e}
		\sqrt{A_{\mathrm E}A_{\mathrm G}}
	}.
	\label{eq:plasma_inductance}
\end{gather}
The integration length is thus approximated by the electrode separation $d_{\mathrm{gap}}$. The quality of this approximation depends on the finite sheath extension and is examined for the reference case below.

The inertial term in Eq.~\eqref{eq:bulk_voltage} gives the inductive bulk response, while $R_{\mathrm p}$ accounts for power dissipation. For purely collisional momentum transfer,
\begin{gather}
	\nu_{\mathrm{eff}}
	=
	\nu_{\mathrm m}.
\end{gather}
Collisionless sheath heating is represented phenomenologically by an additional effective damping term based on the hard-wall description of stochastic heating \cite{Godyak1972},
\begin{gather}
	\nu_{\mathrm{eff}}
	=
	\nu_{\mathrm m}
	+
	\frac{\bar{v}_{\mathrm e}}{d_{\mathrm{gap}}}.
	\label{eq:effective_collision_frequency}
\end{gather}
The second term maps collisionless power absorption onto an effective resistance and should not be interpreted as a microscopic momentum-transfer collision frequency \cite{LiebermanGodyak1998,Mussenbrock2008PRL,Lieberman2008POP}.

\paragraph{Blocking capacitor and self-bias voltage.}

\begin{figure}[t]
	\centering
	\resizebox{\columnwidth}{!}{%
		\begin{circuitikz}[
			american,
			font=\small,
			line width=0.6pt
			]
			
			\ctikzset{
				bipoles/thickness=1
			}
			
			% ------------------------------------------------------------
			% Coordinates
			% ------------------------------------------------------------
			
			\coordinate (GND) at (0,0);
			\coordinate (SRCB) at (0,1.6);
			\coordinate (SRCT) at (0,5.7);
			
			\coordinate (CBL) at (1.5,5.7);
			\coordinate (CBR) at (3.1,5.7);
			
			\coordinate (SHTL) at (4.0,5.7);
			\coordinate (SHTR) at (6.8,5.7);
			\coordinate (SHBL) at (4.0,4.1);
			\coordinate (SHBR) at (6.8,4.1);
			
			\coordinate (SHB) at (5.45,4.1);
			\coordinate (LPB) at (5.45,2.6);
			\coordinate (RPB) at (5.45,1.5);
			\coordinate (VFLB) at (5.45,0.0);
			
			% ------------------------------------------------------------
			% RF source and return
			% ------------------------------------------------------------
			
			\draw
			(GND)
			node[ground]{}
			to[short] (SRCB)
			to[sV,l_={$V_{\mathrm{RF}}(t)$}] (SRCT)
			to[short] (CBL);
			
			% ------------------------------------------------------------
			% Blocking capacitor
			% ------------------------------------------------------------
			
			\draw
			(CBL)
			to[C,l^={$C_{\mathrm B}$}] (CBR)
			to[short] (SHTL);
			
			% Capacitor polarity and voltage label
			\node at (1.75,6.05) {$+$};
			\node at (2.85,6.05) {$-$};
			\node at (2.30,5.00) {$V_{\mathrm B}(t)$};
			
			% ------------------------------------------------------------
			% Current arrow
			% ------------------------------------------------------------
			
			\draw[->]
			(3.5,6.20)
			--
			(4.7,6.20);
			
			\node at (4.1,6.55) {$I(t)$};
			
			% ------------------------------------------------------------
			% Powered sheath block
			% ------------------------------------------------------------
			
			\draw
			(SHTL)
			--
			(SHTR);
			
			\draw
			(SHBL)
			--
			(SHBR);
			
			% Ion current branch
			\draw
			(SHBL)
			to[isource,l_={$I_{\mathrm i}$}]
			(SHTL);
			
			% Electron current branch
			\draw
			(5.45,5.7)
			to[D]
			(5.45,4.1);
			
			\draw[->]
			(5.45,4.5)
			--
			(5.45,4.3);
			
			\node[right] at (5.55,4.3)
			{$I_{\mathrm e}(V_{\mathrm S})$};
			
			% Nonlinear sheath capacitance branch
			\draw
			(SHTR)
			to[C]
			(SHBR);
			
			\node[right] at (6.9,5.20) {$-$};
			\node[right] at (6.9,4.55) {$+$};
			
			\node[right] at (7.3,4.88)
			{$V_{\mathrm S}(t)$};
			
			% ------------------------------------------------------------
			% Plasma bulk
			% ------------------------------------------------------------
			
			\draw
			(SHB)
			to[L,l_={$L_{\mathrm p}$}]
			(LPB)
			to[R,l_={$R_{\mathrm p}$}]
			(RPB);
			
			\node[right] at (6.1,3.45) {$+$};
			\node[right] at (6.1,1.85) {$-$};
			\node[right] at (6.1,2.65) {$V_{\mathrm p}(t)$};
			
			% ------------------------------------------------------------
			% Grounded sheath
			% ------------------------------------------------------------
			
			\draw
			(RPB)
			to[battery1,l_={$V_{\mathrm{fl}}$}]
			(VFLB)
			to[short]
			(GND);
			
			% ------------------------------------------------------------
			% Braces and labels
			% ------------------------------------------------------------
			
			\draw[
			decorate,
			decoration={
				brace,
				amplitude=5pt
			}
			]
			(8.2,5.75)
			--
			(8.2,4.05);
			
			\node[
			right,
			align=center
			] at (8.4,4.90)
			{powered\\sheath};
			
			\draw[
			decorate,
			decoration={
				brace,
				amplitude=5pt
			}
			]
			(8.2,4.0)
			--
			(8.2,1.45);
			
			\node[
			right,
			align=center
			] at (8.4,2.70)
			{plasma\\bulk};
			
			\draw[
			decorate,
			decoration={
				brace,
				amplitude=5pt
			}
			]
			(8.2,1.40)
			--
			(8.2,0.45);
			
			\node[
			right,
			align=center
			] at (8.4,0.92)
			{grounded\\sheath};
			
		\end{circuitikz}
	}
\caption{
	Equivalent circuit of the asymmetric CCRF discharge. The powered sheath contains the nonlinear charge-storage element and the ion and electron conduction currents. The plasma bulk is represented by $L_{\mathrm p}$ and $R_{\mathrm p}$, while the grounded sheath contributes the stationary voltage drop $V_{\mathrm{fl}}$.
}
	\label{fig:equivalent_circuit}
\end{figure}
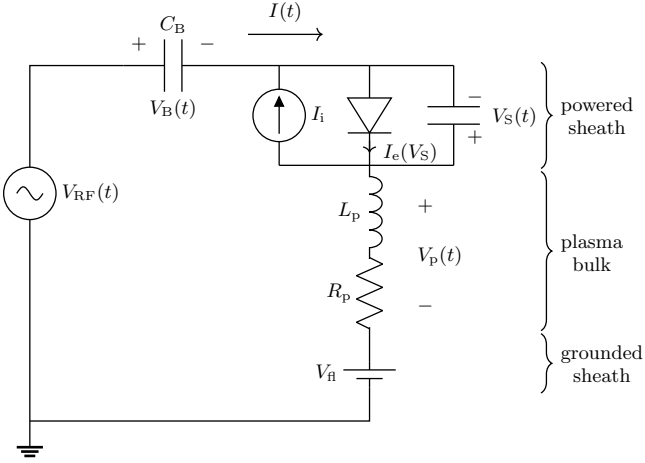

The powered electrode is connected to the RF source through a blocking capacitor $C_{\mathrm B}$, which permits a dc self-bias to develop in the geometrically asymmetric discharge. With the polarity indicated in Fig.~\ref{fig:equivalent_circuit},
\begin{gather}
	V_{\mathrm B}
	=
	\phi_{\mathrm{RF}}
	-
	\phi_{\mathrm E},
	\label{eq:blocking_voltage_definition}
\end{gather}
where
\begin{gather}
	\phi_{\mathrm{RF}}
	=
	V_{\mathrm{RF}}(t).
\end{gather}
The capacitor current is
\begin{gather}
	I(t)
	=
	C_{\mathrm B}
	\frac{dV_{\mathrm B}}{dt}.
	\label{eq:blocking_capacitor}
\end{gather}

For a sinusoidal source,
\begin{gather}
	\left\langle
	V_{\mathrm{RF}}
	\right\rangle_{\mathrm{RF}}
	=
	0 .
\end{gather}
The dc self-bias is the time-averaged powered-electrode potential,
\begin{gather}
	V_{\mathrm{SB}}
	=
	\left\langle
	\phi_{\mathrm E}
	\right\rangle_{\mathrm{RF}}
	=
	-
	\left\langle
	V_{\mathrm B}
	\right\rangle_{\mathrm{RF}} .
	\label{eq:self_bias_average}
\end{gather}
In the stationary periodic state,
\begin{gather}
	\left\langle
	I(t)
	\right\rangle_{\mathrm{RF}}
	=
	0 .
	\label{eq:zero_dc_current}
\end{gather}

The resulting equivalent circuit is shown in Fig.~\ref{fig:equivalent_circuit}.

\subsection{Dynamical equations}

The fast RF dynamics are described by three variables: the powered-sheath charge $Q_{\mathrm S}(t)$, the blocking-capacitor voltage $V_{\mathrm B}(t)$, and the total discharge current $I(t)$. The applied voltage is taken to be sinusoidal,
\begin{gather}
	V_{\mathrm{RF}}(t)
	=
	\widehat{V}_{\mathrm{RF}}
	\cos\left(
	\omega_{\mathrm{RF}}t
	\right).
	\label{eq:rf_voltage}
\end{gather}

With the current directions defined in Fig.~\ref{fig:equivalent_circuit}, charge conservation at the powered sheath gives
\begin{gather}
	\frac{dQ_{\mathrm S}}{dt}
	=
	I_{\mathrm e}(V_{\mathrm S})
	-
	I_{\mathrm i}
	-
	I .
	\label{eq:dynamical_sheath}
\end{gather}

The blocking-capacitor voltage obeys
\begin{gather}
	\frac{dV_{\mathrm B}}{dt}
	=
	\frac{I}{C_{\mathrm B}}.
	\label{eq:dynamical_bias}
\end{gather}

Kirchhoff's voltage law gives
\begin{gather}
	V_{\mathrm{RF}}
	=
	V_{\mathrm B}
	-
	V_{\mathrm S}
	+
	V_{\mathrm p}
	+
	V_{\mathrm{fl}} .
	\label{eq:kirchhoff_voltage}
\end{gather}
Using Eq.~\eqref{eq:bulk_voltage}, the current therefore satisfies
\begin{gather}
	\frac{dI}{dt}
	=
	\frac{1}{L_{\mathrm p}}
	\left[
	V_{\mathrm{RF}}(t)
	-
	V_{\mathrm B}
	+
	V_{\mathrm S}(Q_{\mathrm S})
	-
	V_{\mathrm{fl}}
	-
	R_{\mathrm p}I
	\right].
	\label{eq:dynamical_current}
\end{gather}

Equations~\eqref{eq:dynamical_sheath}, \eqref{eq:dynamical_bias}, and \eqref{eq:dynamical_current} form a closed three-dimensional nonlinear system for a prescribed plasma state $(n_{\mathrm e},n_{\mathrm s},T_{\mathrm e})$. The plasma state fixes the particle currents, sheath charge-voltage relation, bulk impedance, and grounded-sheath voltage.

The nonlinear sheath generates harmonics of the applied excitation, which can couple to the inertial bulk response and excite the plasma series resonance \cite{MussenbrockBrinkmann2006APL,Mussenbrock2008PRL,Lieberman2008POP}. In the self-consistent model developed below, the plasma state is determined from the stationary particle and energy balances together with the cycle-averaged power obtained from the periodic RF solution.

\subsection{Stationary zero-dimensional plasma closure}

The RF model determines the periodic electrical response for a given macroscopic plasma state. To close the model, the electron density and temperature are obtained from stationary particle and energy balances. Both quantities are assumed to remain constant during one RF period, consistent with the separation of the RF and plasma-sustainment timescales introduced above.

We restrict the closure to an electropositive plasma with one dominant positive ion species, so that
\begin{gather}
	n_{\mathrm i}
	\simeq
	n_{\mathrm e}.
	\label{eq:electropositive_quasineutrality}
\end{gather}
The electron density $n_{\mathrm e}$ represents the quasineutral bulk density used in the zero-dimensional balances. The density at the sheath edge is written as
\begin{gather}
	n_{\mathrm s}
	=
	h n_{\mathrm e},
	\label{eq:sheath_edge_density}
\end{gather}
where $h$ accounts for the density reduction toward the sheath boundary. Electronegative plasmas would require additional particle balances and modified sheath-edge conditions but would not alter the structure of the fast RF subsystem.

Throughout the following, $T_{\mathrm e}$ and electron energy losses are expressed in electronvolts, while $T_{\mathrm g}$ is expressed in kelvin. The elementary charge $e$ is included explicitly when quantities given in electronvolts are converted to SI energy units.

\paragraph{Particle balance.}

For a neutral-gas pressure $p$ and temperature $T_{\mathrm g}$,
\begin{gather}
	n_{\mathrm g}
	=
	\frac{p}{k_{\mathrm B}T_{\mathrm g}}.
	\label{eq:neutral_density}
\end{gather}
Electron-impact ionization produces charged particles at the rate
\begin{gather}
	\Gamma_{\mathrm{iz}}
	=
	n_{\mathrm e}
	n_{\mathrm g}
	K_{\mathrm{iz}}(T_{\mathrm e})
	V,
	\label{eq:total_ionization_rate}
\end{gather}
where $V$ is the effective plasma volume and $K_{\mathrm{iz}}$ the ionization rate coefficient. Particle losses are represented by the Bohm flux through an effective loss area $A_{\mathrm{loss}}$,
\begin{gather}
	\Gamma_{\mathrm{wall}}
	=
	n_{\mathrm s}
	v_{\mathrm B}
	A_{\mathrm{loss}},
	\label{eq:wall_loss_rate}
\end{gather}
with
\begin{gather}
	v_{\mathrm B}
	=
	\sqrt{
		\frac{eT_{\mathrm e}}{m_{\mathrm i}}
	}.
	\label{eq:bohm_velocity}
\end{gather}

The stationary condition $\Gamma_{\mathrm{iz}}=\Gamma_{\mathrm{wall}}$, together with $n_{\mathrm s}=h n_{\mathrm e}$, gives
\begin{gather}
	n_{\mathrm g}
	K_{\mathrm{iz}}(T_{\mathrm e})
	=
	h
	v_{\mathrm B}(T_{\mathrm e})
	\frac{A_{\mathrm{loss}}}{V}.
	\label{eq:temperature_balance}
\end{gather}
The plasma density cancels from this equation. For given pressure, gas temperature, geometry, ion species, and sheath-edge factor, Eq.~\eqref{eq:temperature_balance} therefore determines the stationary electron temperature.

\paragraph{Energy balance.}

The nonlinear RF solution provides the cycle-averaged power transferred to the electrons. Since dissipation in the reduced circuit is represented by $R_{\mathrm p}$,
\begin{gather}
	P_{\mathrm{abs}}
	=
	R_{\mathrm p}
	\left\langle
	I^2
	\right\rangle_{\mathrm{RF}}
	=
	L_{\mathrm p}\nu_{\mathrm{eff}}
	\left\langle
	I^2
	\right\rangle_{\mathrm{RF}}.
	\label{eq:absorbed_power_mean_square_current}
\end{gather}

The stationary electron energy balance requires
\begin{gather}
	P_{\mathrm{abs}}
	=
	P_{\mathrm{loss}}.
	\label{eq:global_energy_balance}
\end{gather}
We describe the losses by an effective energy
$\mathcal{E}_{\mathrm{eff}}(T_{\mathrm e})$ per electron--ion pair created and lost from the discharge. This quantity includes collisional energy losses and energy carried by electrons lost to the boundaries. Energy gained by positive ions in the boundary sheaths is not included because $P_{\mathrm{abs}}$ denotes power transferred to the electron population.

The total electron energy-loss rate is
\begin{gather}
	P_{\mathrm{loss}}
	=
	e\,
	n_{\mathrm e}
	n_{\mathrm g}
	K_{\mathrm{iz}}(T_{\mathrm e})
	V
	\mathcal{E}_{\mathrm{eff}}(T_{\mathrm e}).
	\label{eq:global_power_loss}
\end{gather}
Using the particle balance,
\begin{gather}
	P_{\mathrm{loss}}
	=
	e\,
	h n_{\mathrm e}
	v_{\mathrm B}
	A_{\mathrm{loss}}
	\mathcal{E}_{\mathrm{eff}}(T_{\mathrm e}),
	\label{eq:global_power_loss_wall}
\end{gather}
and hence
\begin{gather}
	n_{\mathrm e}
	=
	\frac{P_{\mathrm{abs}}}
	{
		e\,
		h
		v_{\mathrm B}
		A_{\mathrm{loss}}
		\mathcal{E}_{\mathrm{eff}}(T_{\mathrm e})
	}.
	\label{eq:stationary_electron_density}
\end{gather}
Thus, the particle balance fixes $T_{\mathrm e}$, while the energy balance determines $n_{\mathrm e}$ from the RF power absorption.

\section{Model implementation for argon}

The general formulation developed above is independent of the working gas. Its specialization to argon requires the corresponding kinetic rate coefficients, the effective electron energy loss, and the sheath-edge density factor. These quantities are specified below using reduced Maxwellian rate-coefficient representations appropriate for the temperature range considered here. Where more detailed cross-section data are available, they provide the physical reference, while the reduced model uses compact analytical fits in order to retain its algebraic structure.

\subsection{Kinetic and transport coefficients}

For an arbitrary electron energy distribution function $f_E(E)$, normalized according to
\begin{gather}
	\int_0^\infty
	f_E(E)\,dE
	=
	1,
\end{gather}
the rate coefficient associated with an electron-neutral process $j$ is
\begin{gather}
	K_j
	=
	\int_0^\infty
	\sigma_j(E)
	v(E)
	f_E(E)
	\,dE,
	\label{eq:general_rate_coefficient}
\end{gather}
where
\begin{gather}
	v(E)
	=
	\sqrt{
		\frac{2eE}{m_{\mathrm e}}
	}.
\end{gather}

For a Maxwellian electron energy distribution,
\begin{gather}
	f_E(E;T_{\mathrm e})
	=
	\frac{2}{\sqrt{\pi}}
	\frac{\sqrt{E}}{T_{\mathrm e}^{3/2}}
	\exp\left(
	-\frac{E}{T_{\mathrm e}}
	\right),
	\label{eq:maxwellian_eedf}
\end{gather}
where both $E$ and $T_{\mathrm e}$ are expressed in electronvolts. Equation~\eqref{eq:general_rate_coefficient} then becomes
\begin{gather}
	K_j(T_{\mathrm e})
	=
	\sqrt{
		\frac{8e}{\pi m_{\mathrm e}}
	}
	T_{\mathrm e}^{-3/2}
	\int_0^\infty
	\sigma_j(E)
	E
	\exp\left(
	-\frac{E}{T_{\mathrm e}}
	\right)
	\,dE.
	\label{eq:maxwellian_rate_coefficient}
\end{gather}

In the present reduced argon model, the required Maxwellian rate coefficients are represented by analytical fits in order to avoid repeated numerical evaluation of Eq.~\eqref{eq:maxwellian_rate_coefficient} during the self-consistent iteration.

For electron-impact ionization, we use \cite{LiebermanLichtenberg2005}
\begin{gather}
	K_{\mathrm{iz}}(T_{\mathrm e})
	=
	2.34\times10^{-14}
	T_{\mathrm e}^{0.59}
	\exp\left(
	-\frac{17.44}{T_{\mathrm e}}
	\right)
	\quad
	\mathrm{m^3\,s^{-1}},
	\label{eq:argon_ionization_rate}
\end{gather}
where $T_{\mathrm e}$ is expressed in electronvolts. The value $17.44\,\mathrm{eV}$ in the exponential is a parameter of the Maxwellian fit and should not be confused with the argon ionization threshold of $15.76\,\mathrm{eV}$. The fit is used here in the range
\begin{gather}
	1~\mathrm{eV}
	\lesssim
	T_{\mathrm e}
	\lesssim
	7~\mathrm{eV}.
	\label{eq:argon_temperature_range}
\end{gather}

The effective electron-impact excitation rate coefficient is represented by\cite{LiebermanLichtenberg2005}
\begin{gather}
	K_{\mathrm{ex}}(T_{\mathrm e})
	=
	2.48\times10^{-14}
	T_{\mathrm e}^{0.33}
	\exp\left(
	-\frac{12.78}{T_{\mathrm e}}
	\right)
	\quad
	\mathrm{m^3\,s^{-1}}.
	\label{eq:argon_excitation_rate}
\end{gather}

For electron-neutral momentum transfer, detailed energy-dependent cross-section data are available from the Phelps argon data set distributed through LXCat \cite{YamabeBuckmanPhelps1983,PitchfordEtAl2017}. For the reduced implementation, however, we use the analytical Maxwellian rate-coefficient fit given by Lieberman and Lichtenberg \cite{LiebermanLichtenberg2005},
\begin{multline}
	K_{\mathrm m}(T_{\mathrm e})
	=
	2.336\times10^{-14}
	T_{\mathrm e}^{1.609}\dots \\ \dots \times
	\exp\left[
	0.0618
	\left(
	\ln T_{\mathrm e}
	\right)^2
	-
	0.1171
	\left(
	\ln T_{\mathrm e}
	\right)^3
	\right]
	\quad
	\mathrm{m^3\,s^{-1}}.
	\label{eq:argon_momentum_rate}
\end{multline}
The corresponding electron-neutral momentum-transfer frequency is
\begin{gather}
	\nu_{\mathrm m}
	=
	n_{\mathrm g}
	K_{\mathrm m}(T_{\mathrm e}).
	\label{eq:argon_momentum_frequency}
\end{gather}

For the elastic electron-neutral contribution to the stationary electron energy balance, we retain a separate coefficient $K_{\mathrm{el}}(T_{\mathrm e})$. In the present implementation, the same analytical fit is used,
\begin{gather}
	K_{\mathrm{el}}(T_{\mathrm e})
	=
	K_{\mathrm m}(T_{\mathrm e}).
	\label{eq:argon_elastic_energy_rate}
\end{gather}
The two symbols are kept distinct because they enter different parts of the reduced model. The coefficient $K_{\mathrm m}$ determines momentum relaxation in the RF bulk response, whereas $K_{\mathrm{el}}$ enters the elastic contribution to the electron energy-loss function. This distinction also permits the two coefficients to be replaced independently by more detailed kinetic data if required.

\subsection{Energy-loss function}

The stationary electron energy balance requires the effective energy loss $\mathcal{E}_{\mathrm{eff}}(T_{\mathrm e})$ per electron--ion pair created and lost from the discharge. For the present argon model, it accounts for ionization, excitation, elastic energy transfer to the neutral gas, and the kinetic energy carried by electrons lost to the boundaries \cite{LiebermanLichtenberg2005}.

The collisional contribution is written as
\begin{gather}
	\mathcal{E}_{\mathrm c}(T_{\mathrm e})
	=
	E_{\mathrm{iz}}
	+
	\frac{
		K_{\mathrm{ex}}(T_{\mathrm e})
	}{
		K_{\mathrm{iz}}(T_{\mathrm e})
	}
	E_{\mathrm{ex}}
	+
	\frac{
		K_{\mathrm{el}}(T_{\mathrm e})
	}{
		K_{\mathrm{iz}}(T_{\mathrm e})
	}
	\frac{3m_{\mathrm e}}{m_{\mathrm i}}
	T_{\mathrm e}.
	\label{eq:argon_collisional_energy_loss}
\end{gather}
Here,
\begin{gather}
	E_{\mathrm{iz}}
	=
	15.76~\mathrm{eV},
	\qquad
	E_{\mathrm{ex}}
	=
	12.14~\mathrm{eV},
	\label{eq:argon_threshold_energies}
\end{gather}
where $E_{\mathrm{ex}}$ represents the effective excitation energy used in the reduced argon chemistry. The three terms in Eq.~\eqref{eq:argon_collisional_energy_loss} describe the energy required for ionization, excitation losses normalized to the ionization rate, and elastic transfer of electron energy to the neutral gas. The neutral-gas thermal energy is neglected in the last term.

For a Maxwellian electron population, electrons lost to a boundary carry an average kinetic energy
\begin{gather}
	\mathcal{E}_{\mathrm{e,wall}}
	=
	2T_{\mathrm e}.
	\label{eq:argon_electron_wall_energy}
\end{gather}
The effective electron energy loss entering the global energy balance is therefore
\begin{gather}
	\mathcal{E}_{\mathrm{eff}}(T_{\mathrm e})
	=
	E_{\mathrm{iz}}
	+
	\frac{
		K_{\mathrm{ex}}(T_{\mathrm e})
	}{
		K_{\mathrm{iz}}(T_{\mathrm e})
	}
	E_{\mathrm{ex}}
	+
	\frac{
		K_{\mathrm{el}}(T_{\mathrm e})
	}{
		K_{\mathrm{iz}}(T_{\mathrm e})
	}
	\frac{3m_{\mathrm e}}{m_{\mathrm i}}
	T_{\mathrm e}
	+
	2T_{\mathrm e}.
	\label{eq:argon_effective_energy_loss}
\end{gather}

Since $P_{\mathrm{abs}}$ in the present model denotes power transferred to the electron population, energy acquired by positive ions during acceleration through the boundary sheaths is not included in $\mathcal{E}_{\mathrm{eff}}$.

\subsection{Sheath-edge density factor}

The sheath-edge density is related to the characteristic bulk density by
\begin{gather}
	n_{\mathrm s}
	=
	h n_{\mathrm e}.
	\label{eq:argon_sheath_edge_density}
\end{gather}
For the present argon model, the reduction factor $h$ is approximated by the standard one-dimensional expression
\begin{gather}
	h
	=
	0.86
	\left(
	3
	+
	\frac{d_{\mathrm{gap}}}
	{2\lambda_{\mathrm i}}
	\right)^{-1/2},
	\label{eq:argon_h_factor}
\end{gather}
commonly used in global models of electropositive discharges \cite{LiebermanLichtenberg2005}. The ion-neutral mean free path is
\begin{gather}
	\lambda_{\mathrm i}
	=
	\frac{1}
	{n_{\mathrm g}\sigma_{\mathrm i}},
	\label{eq:argon_ion_mean_free_path}
\end{gather}
where $\sigma_{\mathrm i}$ is an effective ion-neutral collision cross section.

For the low ion energies relevant here, we use the representative value
\begin{gather}
	\sigma_{\mathrm i}
	=
	1.0\times10^{-18}\ \mathrm{m^2},
	\label{eq:argon_ion_cross_section}
\end{gather}
based on the $\mathrm{Ar}^+$--Ar collision data compiled by Phelps \cite{Phelps1991}.

The factor $h$ therefore introduces the pressure dependence of the sheath-edge density through the ion-neutral mean free path. Together with the neutral density and the ionization coefficient, it enters the stationary particle balance,
\begin{gather}
	n_{\mathrm g}
	K_{\mathrm{iz}}(T_{\mathrm e})
	=
	h
	v_{\mathrm B}(T_{\mathrm e})
	\frac{A_{\mathrm{loss}}}{V}.
	\label{eq:argon_temperature_closure}
\end{gather}
For given pressure, gas temperature, and geometry, this equation determines the stationary electron temperature.

\subsection{Reduced algebraic closure for argon}

The stationary particle balance can be written in the form
\begin{gather}
	\frac{
		K_{\mathrm{iz}}(T_{\mathrm e})
	}{
		v_{\mathrm B}(T_{\mathrm e})
	}
	=
	\frac{1}
	{n_{\mathrm g}d_{\mathrm{eff}}},
	\label{eq:argon_effective_length_balance}
\end{gather}
where
\begin{gather}
	d_{\mathrm{eff}}
	=
	\frac{V}
	{hA_{\mathrm{loss}}}
	\label{eq:argon_effective_loss_length}
\end{gather}
defines an effective particle-loss length of the zero-dimensional model. It is not a geometrical electrode spacing, but combines the plasma volume, the effective particle-loss area, and the sheath-edge density reduction into a single characteristic length. With the approximation for $h$ adopted above, $d_{\mathrm{eff}}$ depends on pressure, neutral-gas temperature, and geometry, but not on the plasma density or electron temperature.

For the argon ionization coefficient of Eq.~\eqref{eq:argon_ionization_rate}, the solution of Eq.~\eqref{eq:argon_effective_length_balance} can be represented over the temperature range relevant here by
\begin{gather}
	T_{\mathrm e}
	\simeq
	\frac{1~\mathrm{eV}}
	{
		0.167
		+
		0.1315
		\log_{10}
		\left[
		\frac{
			n_{\mathrm g}d_{\mathrm{eff}}
		}{
			10^{18}\,\mathrm{m^{-2}}
		}
		\right]
	}.
	\label{eq:argon_algebraic_temperature}
\end{gather}
The argument of the logarithm is dimensionless. The expression makes explicit the weak, approximately logarithmic dependence of the stationary electron temperature on neutral density and effective loss length. For the reference geometry and the pressure range $1$--$50\,\mathrm{Pa}$ considered below, it reproduces the direct numerical solution of the particle balance to within approximately $1.6\,\%$.

Equation~\eqref{eq:argon_algebraic_temperature} is used only as an analytic approximation. The numerical results presented below are obtained by solving Eq.~\eqref{eq:argon_effective_length_balance} directly.

\subsection{Reference discharge configuration}

To evaluate the model quantitatively, we consider an asymmetric reference geometry with powered- and grounded-electrode radii
$r_{\mathrm E}=5.0\,\mathrm{cm}$ and
$r_{\mathrm G}=10.0\,\mathrm{cm}$, respectively, and an axial separation
$d_{\mathrm{gap}}=5.0\,\mathrm{cm}$. The corresponding geometrical asymmetry is
\begin{gather}
	\mathcal{A}
	=
	\frac{A_{\mathrm G}}{A_{\mathrm E}}
	=
	\left(
	\frac{r_{\mathrm G}}{r_{\mathrm E}}
	\right)^2
	=
	4.
	\label{eq:reference_asymmetry}
\end{gather}

For the equivalent conical-frustum current channel, the plasma volume is
\begin{gather}
	V
	=
	\frac{d_{\mathrm{gap}}}{3}
	\left(
	A_{\mathrm E}
	+
	\sqrt{A_{\mathrm E}A_{\mathrm G}}
	+
	A_{\mathrm G}
	\right),
	\label{eq:reference_volume}
\end{gather}
with
\begin{align}
	A_{\mathrm E}
	&=
	\pi r_{\mathrm E}^2,
	\\
	A_{\mathrm G}
	&=
	\pi r_{\mathrm G}^2.
\end{align}

The effective particle-loss area entering the stationary global balance is taken as
\begin{gather}
	A_{\mathrm{loss}}
	=
	A_{\mathrm E}
	+
	A_{\mathrm G}.
	\label{eq:reference_loss_area}
\end{gather}
This quantity belongs to the zero-dimensional particle balance and is distinct from the lateral surface area of the equivalent conical-frustum current channel. The electrical current-path geometry and the global particle-loss geometry are therefore represented separately within the reduced model.

Table~\ref{tab:reference_parameters} summarizes the reference geometry and operating conditions. For the prescribed-plasma RF analysis, the electron temperature and density are fixed at
$T_{\mathrm e}=3.01\,\mathrm{eV}$ and
$n_{\mathrm e}=4.93\times10^{14}\,\mathrm{m^{-3}}$, corresponding to the stationary plasma state obtained for the same reference operating conditions. These quantities are treated as prescribed parameters when the fast nonlinear RF subsystem is characterized independently of the stationary plasma closure. The sheath-edge density and transport quantities listed in the table follow from this prescribed state.

\begin{table}[t]
	\centering
	\caption{
		Reference parameters for the prescribed-plasma RF analysis at
		$p=1.0\,\mathrm{Pa}$ and
		$\widehat{V}_{\mathrm{RF}}=250\,\mathrm{V}$.
	}
	\label{tab:reference_parameters}
	\begin{tabular}{l l l}
		\hline
		Quantity & Symbol & Value \\
		\hline
		\multicolumn{3}{l}{Geometric and operating parameters} \\
		Powered-electrode radius
		& $r_{\mathrm E}$
		& $5.0\,\mathrm{cm}$ \\
		Grounded-electrode radius
		& $r_{\mathrm G}$
		& $10.0\,\mathrm{cm}$ \\
		Axial separation
		& $d_{\mathrm{gap}}$
		& $5.0\,\mathrm{cm}$ \\
		Area ratio
		& $A_{\mathrm G}/A_{\mathrm E}$
		& $4$ \\
		Blocking capacitance
		& $C_{\mathrm B}$
		& $10\,\mathrm{nF}$ \\
		RF frequency
		& $f_{\mathrm{RF}}$
		& $13.56\,\mathrm{MHz}$ \\
		RF voltage amplitude
		& $\widehat{V}_{\mathrm{RF}}$
		& $250\,\mathrm{V}$ \\
		Gas pressure
		& $p$
		& $1.0\,\mathrm{Pa}$ \\
		Neutral-gas temperature
		& $T_{\mathrm g}$
		& $300\,\mathrm{K}$ \\
		\hline
		\multicolumn{3}{l}{Prescribed plasma state} \\
		Electron temperature
		& $T_{\mathrm e}$
		& $3.01\,\mathrm{eV}$ \\
		Electron density
		& $n_{\mathrm e}$
		& $4.93\times10^{14}\,\mathrm{m^{-3}}$ \\
		\hline
		\multicolumn{3}{l}{Derived plasma and transport quantities} \\
		Sheath-edge density factor
		& $h$
		& $0.286$ \\
		Sheath-edge density
		& $n_{\mathrm s}$
		& $1.41\times10^{14}\,\mathrm{m^{-3}}$ \\
		Momentum-transfer frequency
		& $\nu_{\mathrm m}$
		& $3.06\times10^{7}\,\mathrm{s^{-1}}$ \\
		Stochastic damping rate
		& $\nu_{\mathrm{stoch}}$
		& $2.32\times10^{7}\,\mathrm{s^{-1}}$ \\
		Effective damping frequency
		& $\nu_{\mathrm{eff}}$
		& $5.39\times10^{7}\,\mathrm{s^{-1}}$ \\
		\hline
	\end{tabular}
\end{table}

\section{Results}

\subsection{Nonlinear RF dynamics for a prescribed plasma state}

We first examine the fast nonlinear RF subsystem for the prescribed reference state of Table~\ref{tab:reference_parameters}. The plasma state is held fixed throughout this subsection so that the nonlinear electrical response can be characterized independently of the stationary plasma closure.

\begin{figure}[t]
	\centering
	\includegraphics[width=0.95\linewidth]{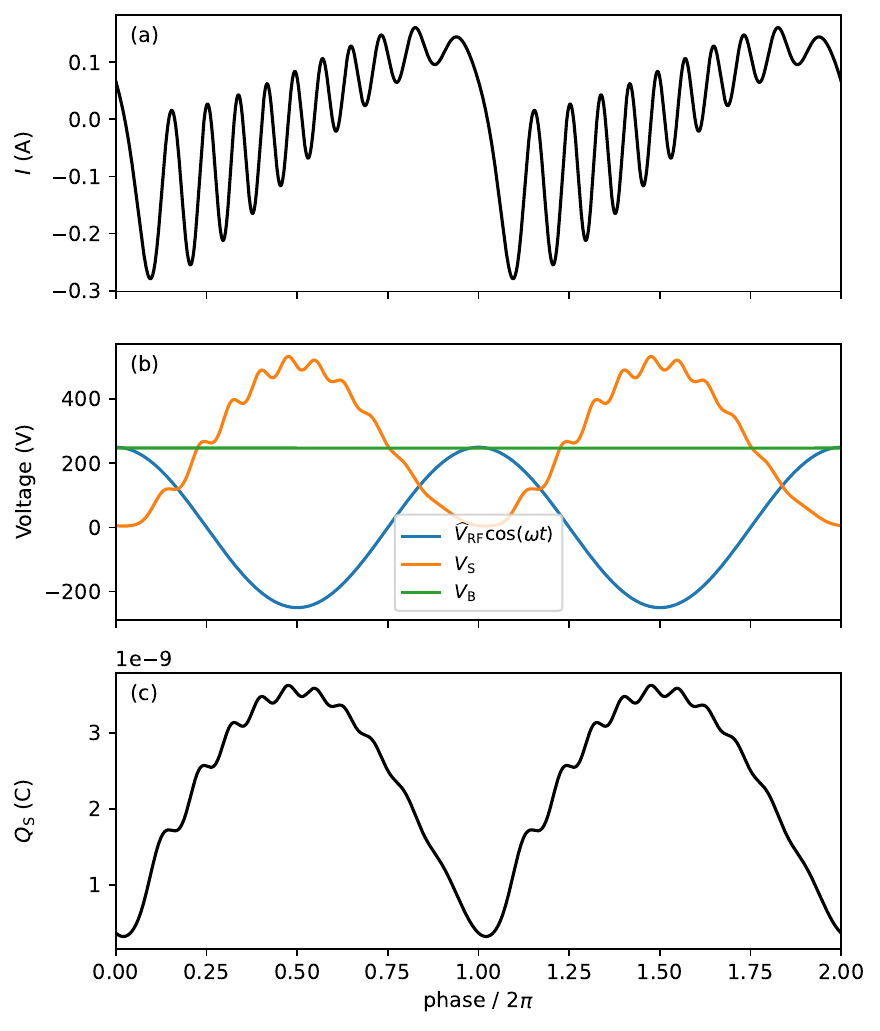}
	\caption{
		Periodic RF solution for the prescribed plasma state at
		$p=1.0\,\mathrm{Pa}$ and
		$\widehat{V}_{\mathrm{RF}}=250\,\mathrm{V}$.
		(a) Discharge current $I(t)$.
		(b) Applied RF voltage, powered-sheath voltage $V_{\mathrm S}$, and blocking-capacitor voltage $V_{\mathrm B}$.
		(c) Powered-sheath charge $Q_{\mathrm S}(t)$.
	}
	\label{fig:rf_dynamics}
\end{figure}

Figure~\ref{fig:rf_dynamics} shows the resulting periodic RF solution. Although the applied voltage is sinusoidal, the discharge current exhibits pronounced high-frequency oscillations. They arise from the nonlinear powered-sheath response interacting with the inertial plasma bulk and are strongest during the rapid variation of the sheath charge. The blocking capacitor develops a finite dc component corresponding to the electrical self-bias. The powered-sheath charge remains non-negative and approaches zero near sheath collapse, consistent with the domain of the quadratic sheath model.

\begin{figure}[t]
	\centering
	\includegraphics[width=0.95\linewidth]{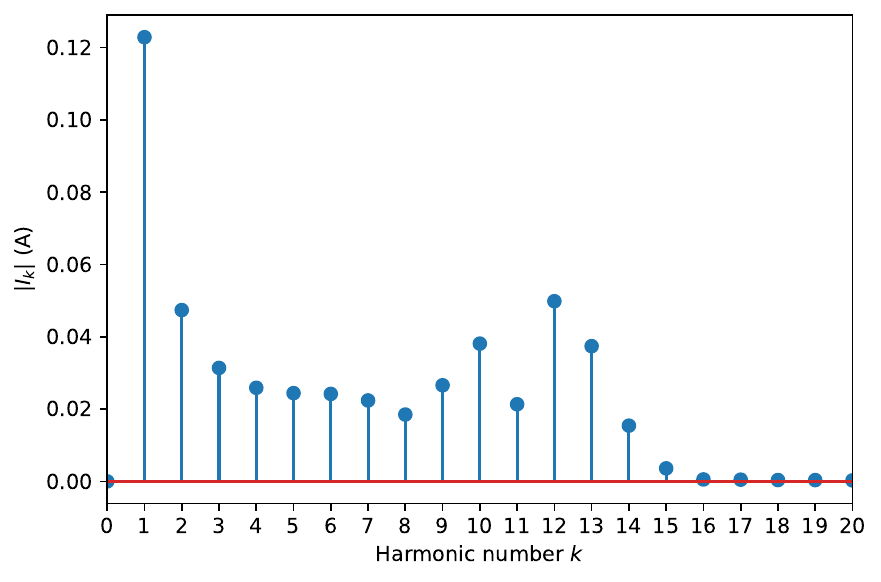}
	\caption{
		Harmonic spectrum $|I_k|$ of the discharge current corresponding to Fig.~\ref{fig:rf_dynamics}. In addition to the fundamental, the nonlinear RF response produces a broad band of enhanced higher harmonics.
	}
	\label{fig:current_spectrum}
\end{figure}

The corresponding current spectrum is shown in Fig.~\ref{fig:current_spectrum}. The higher harmonics do not decay monotonically with harmonic number. Instead, a pronounced band of enhanced components appears approximately between $k=9$ and $k=13$, with the largest amplitude near $k=12$. This structure reflects the coupling of harmonics generated by the nonlinear sheath to the high-frequency response of the plasma bulk. Its relation to the characteristic plasma-series-resonance scale is analyzed below.

\begin{figure}[t]
	\centering
	\includegraphics[width=\columnwidth]{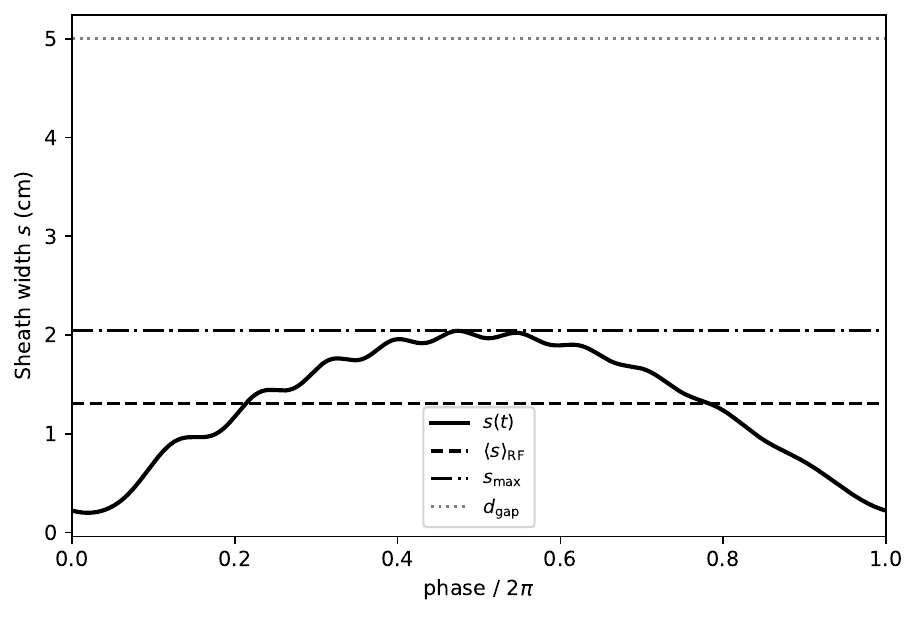}
	\caption{
		Instantaneous powered-sheath width $s(t)$ over one RF period for the prescribed reference case at
		$p=1.0\,\mathrm{Pa}$ and
		$\widehat{V}_{\mathrm{RF}}=250\,\mathrm{V}$.
		The RF-averaged sheath width $\langle s\rangle_{\mathrm{RF}}$, the maximum sheath width $s_{\max}$, and the electrode separation $d_{\mathrm{gap}}$ are shown for comparison.
	}
	\label{fig:sheath_width}
\end{figure}

The powered-sheath voltage also provides an a posteriori estimate of the resolved sheath extension through Eq.~\eqref{eq:sheath_width}. Figure~\ref{fig:sheath_width} shows the resulting sheath width over one RF period. For the reference case,
\begin{align}
	\langle s\rangle_{\mathrm{RF}}
	&=
	1.31\,\mathrm{cm},
	&
	s_{\max}
	&=
	2.04\,\mathrm{cm},
\end{align}
which corresponds to
\begin{align}
	\frac{\langle s\rangle_{\mathrm{RF}}}{d_{\mathrm{gap}}}
	&=
	0.262,
	&
	\frac{s_{\max}}{d_{\mathrm{gap}}}
	&=
	0.408
\end{align}
for $d_{\mathrm{gap}}=5.0\,\mathrm{cm}$.

Figure~\ref{fig:sheath_width} provides an a posteriori consistency check for the characteristic bulk length used in the reduced circuit model. The plasma inductance is evaluated with $d_{\mathrm{gap}}$ as the axial length scale rather than with an explicitly time-dependent bulk length. For the reference discharge, the resolved powered sheath remains smaller than the electrode gap throughout the RF cycle, with an RF-averaged width of about $26\%$ and a maximum width of about $41\%$ of $d_{\mathrm{gap}}$. The approximation is therefore not based on an asymptotic ordering $s\ll d_{\mathrm{gap}}$, but $d_{\mathrm{gap}}$ remains a reasonable characteristic length for the present reduced description.

For smaller gaps or conditions producing substantially wider sheaths, a fixed bulk length would become increasingly restrictive. A more detailed model would then have to account explicitly for the finite and time-dependent sheath extensions in the axial current path.

\subsection{Stationary argon plasma closure}

\begin{figure}[t]
	\centering
	\includegraphics[width=0.85\linewidth]{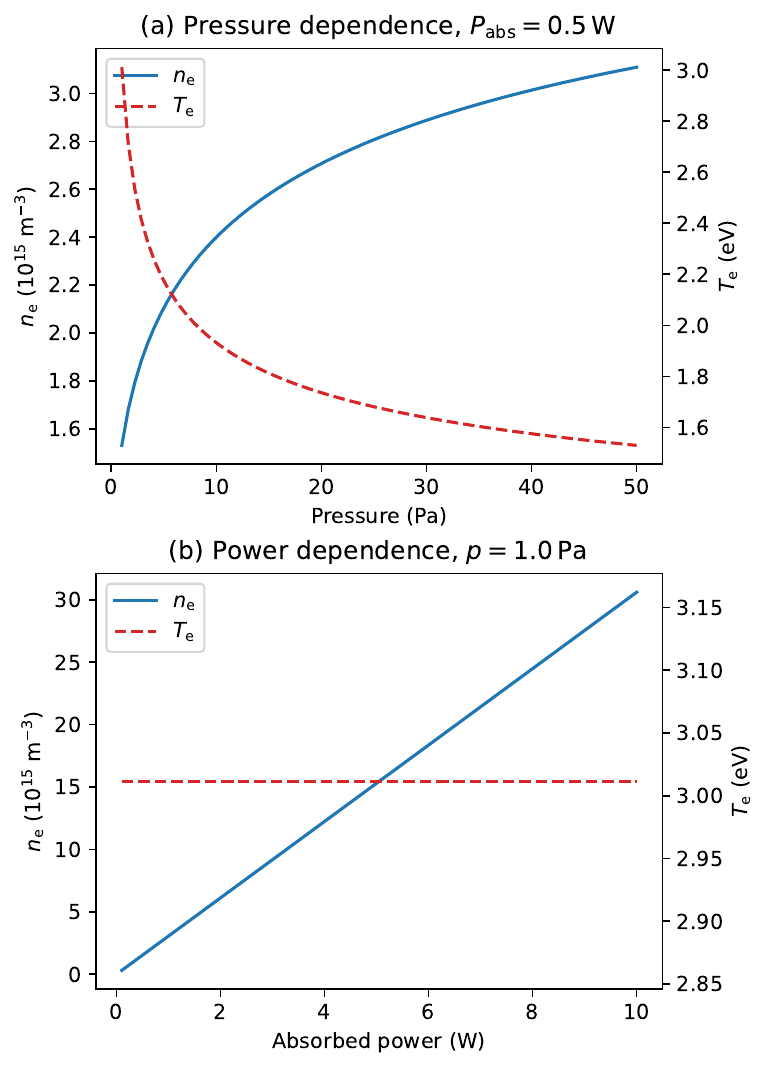}
	\caption{
		Stationary zero-dimensional argon plasma closure.
		(a) Pressure dependence of the electron temperature $T_{\mathrm e}$ and electron density $n_{\mathrm e}$ at fixed absorbed power $P_{\mathrm{abs}}=0.5\,\mathrm{W}$.
		(b) Dependence on absorbed power at fixed pressure $p=1.0\,\mathrm{Pa}$.
	}
	\label{fig:global_closure}
\end{figure}

We next examine the stationary argon closure independently of the nonlinear RF subsystem. Figure~\ref{fig:global_closure} shows the resulting dependence of the electron temperature and density on pressure and absorbed power.

At fixed absorbed power, increasing pressure lowers the stationary electron temperature and raises the electron density. Over the range $1$--$50\,\mathrm{Pa}$ shown in Fig.~\ref{fig:global_closure}(a), $T_{\mathrm e}$ decreases from approximately $3.0$ to $1.55\,\mathrm{eV}$, while $n_{\mathrm e}$ increases from approximately $1.55\times10^{15}$ to $3.1\times10^{15}\,\mathrm{m^{-3}}$. The decrease of $T_{\mathrm e}$ follows from the particle balance: the increase of neutral density and the simultaneous pressure dependence of the sheath-edge factor shift the balance toward lower electron temperature. The corresponding change in the Bohm velocity, sheath-edge factor, and effective electron energy loss leads to the increase of $n_{\mathrm e}$ at fixed absorbed power.

Figure~\ref{fig:global_closure}(b) shows the complementary power dependence at $p=1.0\,\mathrm{Pa}$. Within the present closure, the particle balance is independent of absorbed power, and the stationary electron temperature therefore remains fixed at approximately $3.01\,\mathrm{eV}$. The energy balance then gives
\begin{gather}
	n_{\mathrm e}
	\propto
	P_{\mathrm{abs}},
\end{gather}
which accounts for the linear increase of electron density with absorbed power.

The separation visible in Fig.~\ref{fig:global_closure} is a direct consequence of the adopted stationary closure: the particle balance determines $T_{\mathrm e}$ from pressure and geometry, whereas the absorbed power sets the density through the electron energy balance.

\subsection{Self-consistent coupling and multistart convergence}

\begin{figure}[t]
	\centering
	\begin{tikzpicture}[
		box/.style={
			draw,
			rounded corners,
			align=center,
			minimum width=0.82\columnwidth,
			minimum height=0.75cm
		},
		arrow/.style={
			-{Latex[length=2mm]},
			thick
		},
		node distance=0.85cm
		]
		
		\node[box] (inputs)
		{
			External parameters\\
			$p,\ T_{\mathrm g},\
			\widehat V_{\mathrm{RF}},\
			\omega_{\mathrm{RF}},\
			\text{geometry}$
		};
		
		\node[box,below=of inputs] (te)
		{
			Particle balance\\
			$T_{\mathrm e}$
		};
		
		\node[box,below=of te] (rf)
		{
			Fast nonlinear RF model\\
			$Q_{\mathrm S}(t),\
			V_{\mathrm B}(t),\
			I(t)$
		};
		
		\node[box,below=of rf] (power)
		{
			RF-cycle average\\
			$P_{\mathrm{abs}}
			=
			R_{\mathrm p}
			\langle I^2\rangle_{\mathrm{RF}}$
		};
		
		\node[box,below=of power] (ne)
		{
			Energy balance\\
			$n_{\mathrm e}(P_{\mathrm{abs}})$
		};
		
		\draw[arrow] (inputs) -- (te);
		\draw[arrow] (te) -- (rf);
		\draw[arrow] (rf) -- (power);
		\draw[arrow] (power) -- (ne);
		
		\draw[arrow]
		(ne.east)
		--
		++(0.55,0)
		|-
		(rf.east);
		
	\end{tikzpicture}
	
	\caption{
		Self-consistent coupling between the stationary plasma closure and the fast nonlinear RF model. The particle balance determines $T_{\mathrm e}$, while the RF-cycle-averaged absorbed power determines $n_{\mathrm e}$ through the stationary energy balance. The updated electron density modifies the sheath-edge density and the bulk circuit parameters and thereby closes the feedback loop.
	}
	\label{fig:coupling}
\end{figure}
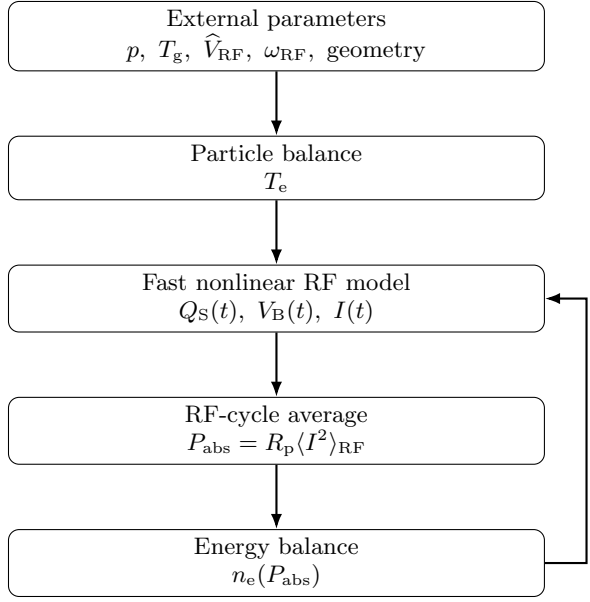

We now couple the nonlinear RF subsystem to the stationary plasma closure. At fixed pressure, neutral-gas temperature, geometry, and RF excitation, the stationary particle balance determines the electron temperature. For a given electron density, the sheath-edge density and the corresponding RF circuit parameters are evaluated, and the nonlinear RF equations are integrated to their periodic state. The resulting cycle-averaged absorbed power provides an updated electron density through the stationary energy balance. The updated density then modifies the RF subsystem and thereby closes the self-consistent feedback loop shown in Fig.~\ref{fig:coupling}.

Neither $T_{\mathrm e}$ nor $n_{\mathrm e}$ is treated as a dynamical variable on the RF timescale. They define the stationary operating point at which the periodic RF solution is evaluated.

To test whether the coupling procedure converges independently of the initial plasma state, we consider three widely separated initial conditions,
\begin{align}
	\left(
	T_{\mathrm e,0},
	n_{\mathrm e,0}
	\right)
	&=
	\left(
	1\,\mathrm{eV},
	10^{14}\,\mathrm{m^{-3}}
	\right),
	\notag\\
	\left(
	T_{\mathrm e,0},
	n_{\mathrm e,0}
	\right)
	&=
	\left(
	3\,\mathrm{eV},
	10^{15}\,\mathrm{m^{-3}}
	\right),
	\notag\\
	\left(
	T_{\mathrm e,0},
	n_{\mathrm e,0}
	\right)
	&=
	\left(
	5\,\mathrm{eV},
	5\times10^{15}\,\mathrm{m^{-3}}
	\right).
	\label{eq:multistart_initial_conditions}
\end{align}
These states serve only as initial conditions and are not required to satisfy the stationary plasma closure.

\begin{figure}[t]
	\centering
	\includegraphics[width=0.85\linewidth]{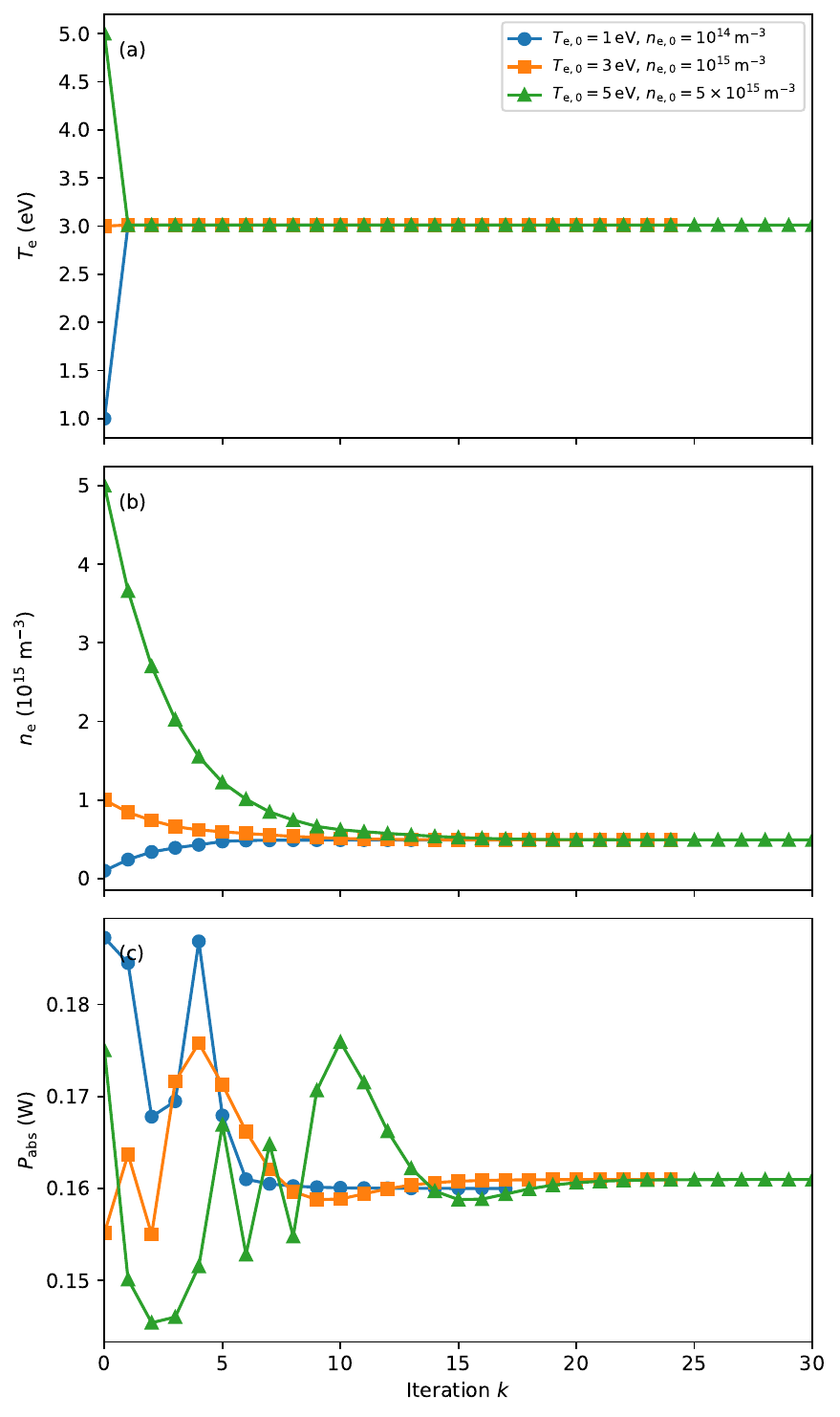}
	\caption{
		Self-consistent coupling iteration for three widely separated initial plasma states at
		$p=1.0\,\mathrm{Pa}$ and
		$\widehat{V}_{\mathrm{RF}}=250\,\mathrm{V}$.
		(a) Electron temperature $T_{\mathrm e}$.
		(b) Electron density $n_{\mathrm e}$.
		(c) Absorbed power $P_{\mathrm{abs}}$.
	}
	\label{fig:multistart_convergence}
\end{figure}

Figure~\ref{fig:multistart_convergence} shows that the three initial electron temperatures are mapped onto the same value of approximately $3.01\,\mathrm{eV}$ after the first coupling update. This follows directly from the stationary particle balance, which fixes $T_{\mathrm e}$ independently of the absorbed power at fixed pressure and geometry.

The electron density and absorbed power exhibit different transient trajectories because the initial densities span almost two orders of magnitude. Nevertheless, all three calculations converge to the same stationary operating point,
\begin{align}
	T_{\mathrm e}
	&\simeq
	3.01\,\mathrm{eV},
	\\
	n_{\mathrm e}
	&\simeq
	4.93\times10^{14}\,\mathrm{m^{-3}},
	\\
	P_{\mathrm{abs}}
	&\simeq
	1.61\times10^{-1}\,\mathrm{W}.
	\label{eq:self_consistent_reference_state}
\end{align}
The multistart calculation therefore indicates a unique converged operating point over the range of initial states considered here.

These self-consistent values are those used as the prescribed plasma state in Table~\ref{tab:reference_parameters}. The RF dynamics discussed above can therefore be interpreted as the fast response evaluated at the self-consistent reference operating point, while the plasma state itself is held fixed during the prescribed-plasma analysis.

\subsection{Self-consistent voltage dependence}

We next use the fully coupled model to determine how the stationary operating point changes with the applied RF-voltage amplitude. The pressure is fixed at
$p=1.0\,\mathrm{Pa}$, while
$\widehat{V}_{\mathrm{RF}}$ is varied from
$100$ to $500\,\mathrm{V}$. At each voltage, the RF subsystem and the stationary plasma closure are iterated to self-consistency.

At fixed pressure and geometry, the stationary particle balance fixes the electron temperature independently of the absorbed power. The electron temperature therefore remains at approximately
$3.01\,\mathrm{eV}$ throughout the voltage scan. The voltage dependence of the stationary state is consequently reflected primarily in the electron density, absorbed power, and discharge current.

At fixed $T_{\mathrm e}$, the effective damping frequency is constant, while
\begin{gather}
	R_{\mathrm p}
	=
	\nu_{\mathrm{eff}}L_{\mathrm p}
	\propto
	\frac{1}{n_{\mathrm e}}.
\end{gather}
The stationary energy balance gives
\begin{gather}
	n_{\mathrm e}
	\propto
	P_{\mathrm{abs}},
\end{gather}
and, together with
$P_{\mathrm{abs}}
=
R_{\mathrm p}
\langle I^2\rangle_{\mathrm{RF}}$,
this implies
\begin{gather}
	\left\langle I^2\right\rangle_{\mathrm{RF}}
	\propto
	n_{\mathrm e}^{\,2}.
	\label{eq:current_density_scaling}
\end{gather}
The three quantities shown below are therefore directly linked by the self-consistent closure.

\begin{figure}[t]
	\centering
	\includegraphics[width=0.85\linewidth]
	{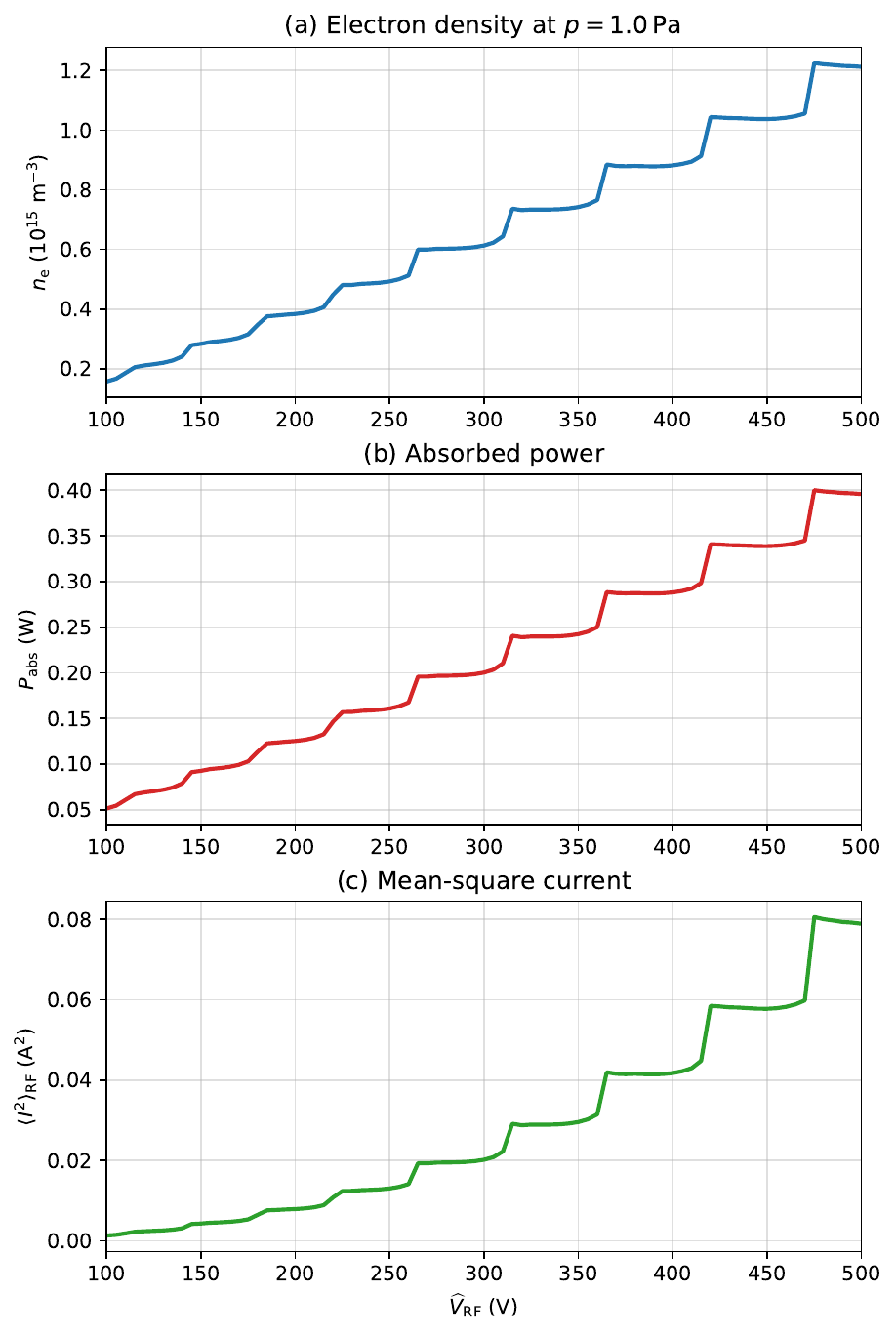}
	\caption{
		Self-consistent voltage scan at $p=1.0\,\mathrm{Pa}$.
		(a) Electron density $n_{\mathrm e}$.
		(b) Absorbed power $P_{\mathrm{abs}}$.
		(c) RF-cycle-averaged mean-square discharge current
		$\langle I^2\rangle_{\mathrm{RF}}$.
	}
	\label{fig:self_consistent_1d_scan}
\end{figure}

Figure~\ref{fig:self_consistent_1d_scan} shows an overall increase of all three quantities with applied voltage. Over the range
$\widehat{V}_{\mathrm{RF}}=100$--$500\,\mathrm{V}$, the electron density rises from approximately
$1.5\times10^{14}$ to
$1.2\times10^{15}\,\mathrm{m^{-3}}$, while the absorbed power increases from about
$0.05$ to
$0.40\,\mathrm{W}$. The mean-square current increases correspondingly by more than an order of magnitude.

The increase is distinctly non-smooth. Pronounced plateaus are separated by relatively sharp transitions, and the same structure appears in
$n_{\mathrm e}$,
$P_{\mathrm{abs}}$, and
$\langle I^2\rangle_{\mathrm{RF}}$.
This correspondence follows from the coupling relations above: changes in the nonlinear RF-current response modify the absorbed power and are transferred directly to the stationary electron density.

The step-like structure indicates that increasing the applied voltage does more than rescale a fixed periodic solution. Instead, the high-frequency current response reorganizes as the self-consistent plasma state changes. As shown in the following subsection, these transitions are closely correlated with the successive shift of the enhanced harmonic band toward higher harmonic numbers.

\subsection{Harmonic excitation and plasma series resonance}

The nonlinear sheath response generates higher harmonics of the discharge current. In the periodic state, the current may be written as
\begin{gather}
	I(t)
	=
	I_0
	+
	\sum_{k=1}^{\infty}
	\widehat{I}_k
	\cos\left(
	k\omega_{\mathrm{RF}}t+\varphi_k
	\right),
	\label{eq:current_fourier}
\end{gather}
where $\widehat{I}_k$ and $\varphi_k$ denote the amplitude and phase of the $k$th harmonic. The resulting discrete spectrum provides a direct measure of the high-frequency response of the nonlinear RF subsystem.

The interpretation of this spectrum in terms of a plasma series resonance requires some care. The powered sheath is nonlinear, and its differential elastance
\begin{gather}
	S_{\mathrm S}(t)
	=
	\frac{\partial V_{\mathrm S}}
	{\partial Q_{\mathrm S}}
\end{gather}
varies strongly during the RF cycle. The periodic state therefore cannot, in general, be represented by a single constant sheath capacitance.

To identify the corresponding high-frequency scale, we consider a small perturbation about the periodic RF solution. On this fast timescale, perturbations of the ion and electron conduction currents are neglected, so that sheath charge conservation gives
\begin{gather}
	\delta I
	\simeq
	-
	\frac{d\,\delta Q_{\mathrm S}}{dt}.
\end{gather}
Linearization of the blocking-capacitor and bulk-current equations then yields
\begin{gather}
	L_{\mathrm p}
	\frac{d^2\delta Q_{\mathrm S}}{dt^2}
	+
	R_{\mathrm p}
	\frac{d\delta Q_{\mathrm S}}{dt}
	+
	\left[
	S_{\mathrm S}(t)
	+
	\frac{1}{C_{\mathrm B}}
	\right]
	\delta Q_{\mathrm S}
	\simeq
	0.
	\label{eq:linearized_psr}
\end{gather}
The high-frequency sheath--bulk dynamics thus correspond to an oscillator with a periodically varying stiffness. Neglecting damping only for the purpose of identifying an instantaneous frequency scale gives
\begin{gather}
	\omega_{\mathrm{loc}}(t)
	=
	\sqrt{
		\frac{
			S_{\mathrm S}(t)+C_{\mathrm B}^{-1}
		}{
			L_{\mathrm p}
		}
	}.
	\label{eq:local_psr_frequency}
\end{gather}
The nonlinear periodic state therefore does not possess a unique time-independent LC resonance frequency.

A useful reference is provided by the classical treatment of the series resonance in bounded RF plasmas \cite{Godyak1986}. In its simplest symmetric planar form, the characteristic frequency is commonly written as
\begin{gather}
	\omega_{\mathrm r}
	\simeq
	\omega_{\mathrm{pe}}
	\sqrt{
		\frac{2\overline{s}}{l_{\mathrm p}}
	},
	\label{eq:godyak_psr}
\end{gather}
where $\overline{s}$ denotes the time-averaged width of one sheath and $l_{\mathrm p}$ the plasma-bulk length. This expression represents the series resonance between the combined sheath capacitance and the inertial response of the bulk electrons.

For the quadratic sheath model used here, the instantaneous sheath charge and width are related by
\begin{gather}
	Q_{\mathrm S}
	=
	e n_{\mathrm s}A_{\mathrm E}s,
\end{gather}
and the differential elastance becomes
\begin{gather}
	S_{\mathrm S}(t)
	=
	\frac{s(t)}
	{\varepsilon_0A_{\mathrm E}}.
	\label{eq:elastance_sheath_width}
\end{gather}
Averaging the sheath width is therefore equivalent, within the present model, to averaging the differential sheath elastance. The correspondence is one of averaging principle rather than circuit identity: the classical expression contains two dynamically participating sheaths, whereas only the powered-electrode sheath is treated dynamically in the present three-variable model.

Moment-based reductions are widely used to extract characteristic scales from fluctuating quantities and distributions, for example in random-vibration and turbulence theory \cite{Rice1944,Vanmarcke1972,TennekesLumley1972}. Motivated by this general construction, we introduce the RF-cycle moments
\begin{gather}
	M_m
	=
	\left\langle
	S_{\mathrm S}^{m}
	\right\rangle_{\mathrm{RF}},
	\qquad
	m=0,1,2,\ldots .
	\label{eq:sheath_moments}
\end{gather}
A characteristic elastance can then be formed from the ratio of two adjacent moments,
\begin{gather}
	S_{\mathrm{eff}}^{(m)}
	=
	\frac{M_{m+1}}{M_m}
	=
	\frac{
		\left\langle
		S_{\mathrm S}^{m+1}
		\right\rangle_{\mathrm{RF}}
	}{
		\left\langle
		S_{\mathrm S}^{m}
		\right\rangle_{\mathrm{RF}}
	}.
	\label{eq:moment_effective_elastance}
\end{gather}
Equivalently,
\begin{gather}
	S_{\mathrm{eff}}^{(m)}
	=
	\left\langle
	w_m(t)S_{\mathrm S}(t)
	\right\rangle_{\mathrm{RF}},
\end{gather}
with
\begin{gather}
	w_m(t)
	=
	\frac{
		S_{\mathrm S}^{m}(t)
	}{
		\left\langle
		S_{\mathrm S}^{m}
		\right\rangle_{\mathrm{RF}}
	}.
\end{gather}
Thus, $S_{\mathrm{eff}}^{(m)}$ is a weighted RF-cycle average of the differential sheath elastance. Increasing the moment order progressively emphasizes those phases of the RF cycle for which $S_{\mathrm S}$ is large.

The resulting hierarchy has a definite ordering. Since $S_{\mathrm S}(t)\geq0$, the moments satisfy
\begin{gather}
	M_{m+1}^{2}
	\leq
	M_mM_{m+2},
	\label{eq:moment_logconvexity}
\end{gather}
as follows from the Cauchy--Schwarz inequality. Dividing by $M_mM_{m+1}>0$ gives
\begin{gather}
	\frac{M_{m+1}}{M_m}
	\leq
	\frac{M_{m+2}}{M_{m+1}},
\end{gather}
and hence
\begin{gather}
	S_{\mathrm{eff}}^{(m)}
	\leq
	S_{\mathrm{eff}}^{(m+1)}.
	\label{eq:moment_ratio_ordering}
\end{gather}
This monotonic ordering is a standard property of moment ratios of non-negative quantities \cite{HardyLittlewoodPolya1952}. For bounded $S_{\mathrm S}(t)$, the hierarchy approaches the maximum elastance,
\begin{gather}
	\lim_{m\rightarrow\infty}
	S_{\mathrm{eff}}^{(m)}
	=
	S_{\max},
	\qquad
	S_{\max}
	=
	\max_t S_{\mathrm S}(t).
	\label{eq:moment_maximum_limit}
\end{gather}
The moment order therefore provides a systematic interpolation between a cycle-averaged description and the instantaneous maximum-elastance limit.

The lowest-order member is
\begin{gather}
	S_{\mathrm{eff}}^{(0)}
	=
	\left\langle
	S_{\mathrm S}
	\right\rangle_{\mathrm{RF}}.
	\label{eq:mean_elastance}
\end{gather}
Through Eq.~\eqref{eq:elastance_sheath_width}, this corresponds directly to averaging the sheath width and therefore provides, within the present one-dynamic-sheath model, the closest analogue of the time-averaged-sheath construction entering the classical resonance estimate.

The second-order member has a particularly direct interpretation for the quadratic sheath model. The sheath voltage and differential elastance are
\begin{gather}
	V_{\mathrm S}
	=
	\frac{Q_{\mathrm S}^{2}}
	{2e\varepsilon_0n_{\mathrm s}A_{\mathrm E}^{2}},
	\qquad
	S_{\mathrm S}
	=
	\frac{Q_{\mathrm S}}
	{e\varepsilon_0n_{\mathrm s}A_{\mathrm E}^{2}}.
\end{gather}
Eliminating $Q_{\mathrm S}$ gives
\begin{gather}
	V_{\mathrm S}
	=
	\frac{
		e\varepsilon_0n_{\mathrm s}A_{\mathrm E}^{2}
	}{2}
	S_{\mathrm S}^{2}.
	\label{eq:voltage_elastance_relation}
\end{gather}
It follows that
\begin{align}
	S_{\mathrm{eff}}^{(2)}
	&=
	\frac{
		\left\langle
		S_{\mathrm S}^{3}
		\right\rangle_{\mathrm{RF}}
	}{
		\left\langle
		S_{\mathrm S}^{2}
		\right\rangle_{\mathrm{RF}}
	}
	\\
	&=
	\frac{
		\left\langle
		V_{\mathrm S}S_{\mathrm S}
		\right\rangle_{\mathrm{RF}}
	}{
		\left\langle
		V_{\mathrm S}
		\right\rangle_{\mathrm{RF}}
	}.
	\label{eq:voltage_weighted_elastance}
\end{align}
The $m=2$ member is therefore a voltage-weighted differential sheath elastance. It preferentially samples the expanded-sheath part of the RF cycle, while remaining a cycle-averaged quantity rather than an instantaneous extremum.

For each moment order, a corresponding characteristic PSR harmonic number is defined by
\begin{gather}
	k_{\mathrm{PSR}}^{(m)}
	=
	\frac{1}{\omega_{\mathrm{RF}}}
	\sqrt{
		\frac{
			S_{\mathrm{eff}}^{(m)}
			+
			C_{\mathrm B}^{-1}
		}{
			L_{\mathrm p}
		}
	}.
	\label{eq:moment_psr}
\end{gather}
The ordinary RF-cycle average is recovered for $m=0$, whereas the maximum-elastance estimate follows in the limit $m\rightarrow\infty$. The second-order member provides an intermediate characteristic scale obtained directly from low-order moments of the nonlinear sheath waveform. Equation~\eqref{eq:moment_psr} is not intended to predict a unique spectral peak, but to characterize the frequency scale of the periodically varying sheath--bulk system.

The corresponding characteristic angular frequency is
\begin{gather}
	\omega_{\mathrm{PSR}}^{(m)}
	=
	\sqrt{
		\frac{
			S_{\mathrm{eff}}^{(m)}
			+
			C_{\mathrm B}^{-1}
		}{
			L_{\mathrm p}
		}
	}.
	\label{eq:moment_psr_frequency}
\end{gather}
Using
\begin{gather}
	L_{\mathrm p}
	=
	\frac{
		m_{\mathrm e}d_{\mathrm{gap}}
	}{
		e^2n_{\mathrm e}\sqrt{A_{\mathrm E}A_{\mathrm G}}
	}
\end{gather}
and
\begin{gather}
	\omega_{\mathrm{pe}}
	=
	\sqrt{
		\frac{
			e^2n_{\mathrm e}
		}{
			m_{\mathrm e}\varepsilon_0
		}
	},
\end{gather}
this expression can be written in the compact form
\begin{gather}
	\omega_{\mathrm{PSR}}^{(m)}
	=
	\omega_{\mathrm{pe}}
	\sqrt{
		\frac{
			C_0
		}{
			C_{\mathrm{eff}}^{(m)}
		}
	},
	\label{eq:psr_plasma_frequency}
\end{gather}
where
\begin{gather}
	C_0
	=
	\frac{
		\varepsilon_0\sqrt{A_{\mathrm E}A_{\mathrm G}}
	}{
		d_{\mathrm{gap}}
	}
\end{gather}
and
\begin{gather}
	\frac{1}{C_{\mathrm{eff}}^{(m)}}
	=
	S_{\mathrm{eff}}^{(m)}
	+
	\frac{1}{C_{\mathrm B}}.
\end{gather}
This representation relates the characteristic PSR scale to the electron plasma frequency while retaining the capacitive contribution of the nonlinear sheath and the blocking capacitor. Since $C_{\mathrm{eff}}^{(m)}$ depends on the self-consistent sheath waveform and plasma state, the expression is a factorization of the coupled response rather than a separation into independent bulk and sheath contributions.

\begin{figure*}[t]
	\centering
	\includegraphics[width=0.9\textwidth]{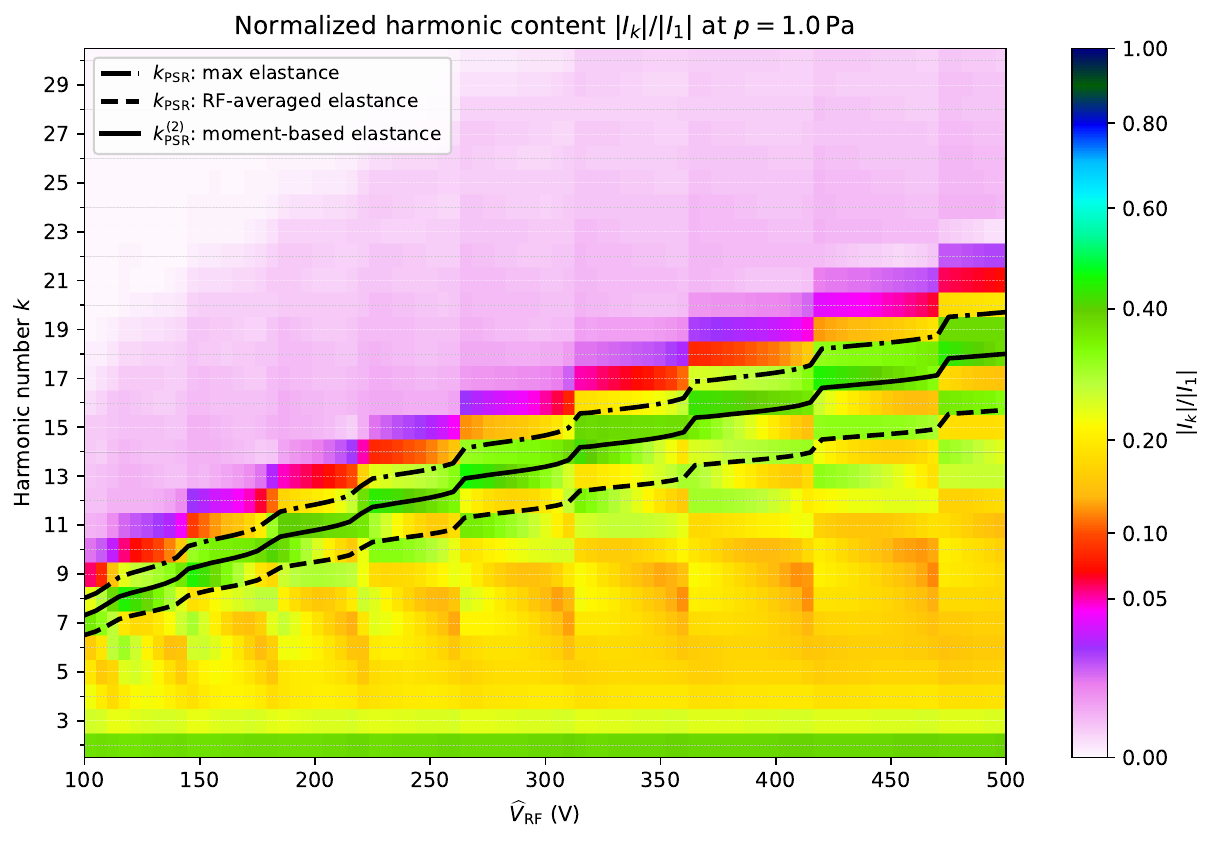}
	\caption{
		Normalized current-harmonic amplitudes
		$\widehat{I}_k/\widehat{I}_1$ as a function of applied RF-voltage amplitude at
		$p=1.0\,\mathrm{Pa}$.
		The curves show characteristic PSR harmonic numbers obtained from the maximum sheath elastance, the ordinary RF-cycle average
		$S_{\mathrm{eff}}^{(0)}
		=
		\langle S_{\mathrm S}\rangle_{\mathrm{RF}}$,
		and the second-order moment
		$S_{\mathrm{eff}}^{(2)}
		=
		\langle S_{\mathrm S}^{3}\rangle_{\mathrm{RF}}/
		\langle S_{\mathrm S}^{2}\rangle_{\mathrm{RF}}$.
		A square-root color normalization is used to resolve weaker higher-order components.
	}
	\label{fig:harmonic_map}
\end{figure*}

Figure~\ref{fig:harmonic_map} shows a broad band of enhanced higher harmonics whose position shifts systematically toward larger harmonic numbers with increasing RF voltage. The three characteristic estimates exhibit the ordering implied by Eq.~\eqref{eq:moment_ratio_ordering}. The RF-averaged elastance gives the lowest characteristic harmonic number and lies systematically below the central part of the enhanced band, whereas the maximum-elastance estimate lies above it. The second-order moment remains between these limits and follows the enhanced harmonic band closely over the complete voltage range.

The comparison also illustrates the physical meaning of the moment hierarchy. The ordinary RF-cycle average gives equal temporal weight to all phases of the RF cycle, including sheath collapse, where the differential elastance is small. The maximum-elastance estimate represents the opposite limit and reduces the cycle to the instant of largest sheath extension. Increasing the moment order continuously shifts the weighting toward the expanded-sheath phase. The second-order member retains information from the complete RF cycle while emphasizing precisely this part of the sheath dynamics.

For the present model, $S_{\mathrm{eff}}^{(2)}$ therefore provides a compact characteristic measure of the PSR scale. Its significance does not follow from fitting the harmonic spectrum: it is the second-order member of the systematically ordered moment hierarchy and, for the quadratic sheath relation, is independently identified as the voltage-weighted differential elastance. The close correspondence with the enhanced harmonic band in Fig.~\ref{fig:harmonic_map} is consequently a result of the comparison rather than an assumption of the construction.

With increasing RF voltage, both the self-consistent plasma density and the sheath waveform change. The plasma inductance and effective sheath elastance therefore vary simultaneously, shifting the characteristic PSR scale through the discrete harmonic spectrum. The enhanced harmonic band follows this shift and undergoes successive reorganizations as the applied voltage increases.

These spectral changes correlate with the plateau-and-transition structure of
$\langle I^2\rangle_{\mathrm{RF}}$,
$P_{\mathrm{abs}}$, and
$n_{\mathrm e}$ in Fig.~\ref{fig:self_consistent_1d_scan}. The coupled model thereby links the nonlinear harmonic response to RF power absorption and, through the stationary energy balance, to the plasma density. We do not assign individual steps to unique harmonic transitions, since the high-frequency response extends over a finite band rather than a single spectral component.

\section{Conclusion and outlook}

We have developed a minimal self-consistent model for the nonlinear RF dynamics of a geometrically asymmetric capacitively coupled plasma. The fast subsystem contains only three dynamical variables: the powered-sheath charge, the blocking-capacitor voltage, and the discharge current. Nonlinear sheath charging, electron inertia, resistive damping, dc self-bias formation, and harmonic generation are retained within this reduced description. Coupling the periodic RF solution to stationary particle and electron-energy balances determines the electron temperature and density without introducing additional dynamical variables on the RF timescale.

For the argon discharge considered here, the particle balance fixes the stationary electron temperature primarily through pressure and geometry, whereas the absorbed RF power determines the electron density. The density in turn modifies the sheath-edge density and the bulk inductance and resistance, thereby closing the feedback between the stationary plasma state and the nonlinear RF response. Multistart calculations converge to the same stationary operating point over the range of initial conditions investigated.

The fully coupled voltage scan reveals a pronounced non-smooth response. As the applied RF voltage is increased, the enhanced harmonic band shifts toward progressively higher harmonic numbers. The corresponding reorganizations of the current spectrum correlate with the plateau-and-transition structure observed in the mean-square current, absorbed power, and stationary electron density. The model therefore connects changes in the nonlinear high-frequency response directly to the macroscopic plasma state through the RF power balance.

A central result concerns the characteristic plasma-series-resonance scale. Linearization about the periodic RF state shows that the high-frequency sheath--bulk dynamics are governed by a periodically varying differential sheath elastance and therefore cannot, in general, be reduced to a unique fixed LC resonance frequency. A hierarchy of effective elastances based on adjacent RF-cycle moments provides a systematic reduction of this time-dependent quantity. For the quadratic sheath model, the second-order member is equivalent to a voltage-weighted differential sheath elastance. It lies between the ordinary RF-cycle average and the maximum-elastance limit and follows the voltage-dependent band of enhanced harmonics substantially more closely than either limiting estimate.

The resulting moment-based quantity should be interpreted as a characteristic PSR scale rather than as the prediction of a unique spectral peak. Its value is obtained directly from the periodic state and does not require a posteriori identification of a harmonic maximum. This makes the construction particularly useful for reduced-order descriptions in which the characteristic high-frequency response is to be related directly to the plasma state.

The present model deliberately retains only the minimum structure required for this coupling. The grounded-electrode sheath is not treated dynamically, the plasma bulk is represented by a fixed characteristic axial length, and the particle and energy balances are stationary and spatially global. Extensions to two dynamical sheaths, finite and time-dependent bulk lengths, and more detailed kinetic or chemical closures are therefore natural next steps. Comparison with kinetic simulations and experiments will be required to determine how far the moment-based PSR description remains valid beyond the present reduced model. Its relation to measurable harmonic currents may ultimately provide a useful basis for plasma-state diagnostics and feedback concepts.

\begin{acknowledgments}
	This work was funded by the Deutsche Forschungsgemeinschaft (DFG, German Research Foundation) under research grant MU 2332/12-1 (project no. 534102992).
\end{acknowledgments}

%\section*{Author Declarations}

%\subsection*{Conflict of Interest}

%The author has no conflicts to disclose.

\section*{Data Availability}

The data that support the findings of this study are available from the corresponding author upon reasonable request.

	\bibliography{M-002-MinimalRFCCP-References}

\end{document}